\documentclass[submission]{SciPost}

\usepackage{graphicx}
\usepackage{dcolumn}
\usepackage{bm}
\usepackage{nicefrac}
\usepackage{xcolor}
\usepackage{hyperref}
\usepackage{gensymb}
\usepackage{amsmath}
\usepackage{xcolor}
\usepackage{etoolbox}
\usepackage{physics}
\usepackage{amssymb}
\usepackage{soul}
\usepackage{subcaption}
\usepackage{float}
\usepackage{booktabs}
\usepackage{tikz} 

\usepackage{orcidlink}

\definecolor{tyrianpurple}{rgb}{0.4, 0.01, 0.24}

\hypersetup{
    colorlinks,
    linkcolor={red!50!black},
    citecolor={blue!50!black},
    urlcolor={blue!80!black}
}

\usepackage[bitstream-charter]{mathdesign}
\usetikzlibrary{arrows.meta, positioning}
\newcommand{\cinam}{Aix-Marseille Univ., CNRS, CINaM, Centre Interdisciplinaire de Nanoscience de Marseille, UMR 7325, Campus de Luminy, 13288 Marseille cedex 9, France}

\newcommand{\NF}{N_F}
\newcommand{\ee}{\mathrm{e}}
\newcommand{\ii}{\mathrm{i}}
\graphicspath{{./}}

\usepackage[english]{babel}
\usepackage[utf8]{inputenc}
\usepackage[T1]{fontenc}
\let\hbar\relax
\DeclareRobustCommand{\hbar}{\mathchar'26\mkern-9mu h}
\hypersetup{hypertexnames=false}

\fancypagestyle{SPstyle}{
\fancyhf{}
\lhead{\colorbox{scipostblue}{\bf \color{white} ~SciPost Physics Codebases }}
\rhead{{\bf \color{scipostdeepblue} {SciPost Phys. Codebases (2026)}}}

\fancyfoot[C]{\textbf{\thepage}}
}

\begin{document}

\pagestyle{SPstyle}

\begin{center}{\Large \textbf{A Julia implementation of Floquet-Bloch theory \\for high-harmonic generation in periodic systems}}\end{center}
\begin{center}
Ashwin Pillai\textsuperscript{1*}, 
Aurelien Manchon\textsuperscript{1}\orcidlink{0000-0002-4768-293X},
Claudio Attaccalite\textsuperscript{1}\orcidlink{0000-0002-7660-261X}
\end{center}


%
\begin{center}
{\bf 1} \cinam
\\
$\star$ \href{mailto:ashwin.pillai@etu.univ-amu.fr}{\small ashwin.pillai@etu.univ-amu.fr}
\end{center}
\date{\today}

\section*{\color{scipostdeepblue}{Abstract}}
\textbf{\boldmath{%
We present a flexible Julia codebase implementing Floquet-Bloch theory for high-harmonic generation (HHG) in periodic tight-binding systems. Using the frequency-domain Fourier-Sambe formalism, the package maps the time-dependent Schrödinger equation onto a static eigenvalue problem. This allows direct computation of Floquet states, quasienergy bands, and harmonic currents for 1D, 2D, or 3D lattices under arbitrary light polarizations. The implementation supports both analytical Fourier-Bessel expansions (via Peierls substitution) and numerical Fast Fourier Transforms (FFT). We validate the code against a canonical 1D two-band model and showcase its capabilities on 2D honeycomb systems: gapped graphene—mapping HHG cutoff scaling with hopping parameters—and the Kane-Mele model, exploring field-induced quasienergy gaps and topological transitions under linear and circular driving fields. In addition to the theoretical presentation, we provide a flexible and efficient open-source Julia implementation of the Floquet formalism: \href{https://github.com/Ashwin-A-Pillai/Floquet-Fourier-Numericals/tree/main}{https://github.com/Ashwin-A-Pillai/Floquet-Fourier-Numericals/tree/main}. }} \vspace{\baselineskip}

	

\noindent\textcolor{white!90!black}{%
\fbox{\parbox{0.975\linewidth}{%
\textcolor{white!40!black}{\begin{tabular}{lr}%
  \begin{minipage}{0.6\textwidth}%
    {\small Copyright attribution to authors. \newline
    This work is a submission to SciPost Physics Codebases. \newline
    License information to appear upon publication. \newline
    Publication information to appear upon publication.}
  \end{minipage} & \begin{minipage}{0.4\textwidth}
    {\small Received Date \newline Accepted Date \newline Published Date}%
  \end{minipage}
\end{tabular}}
}}
}

\vspace{10pt}
\noindent\rule{\textwidth}{1pt}
\tableofcontents
\noindent\rule{\textwidth}{1pt}
\vspace{10pt}

	\section{Introduction}
High-harmonic generation (HHG) is the emission of radiation at integer
	multiples of the frequency of an intense driving field. In weak-field
	nonlinear optics, the response at the \(N\)-th harmonic is described by a
	perturbative susceptibility and decreases rapidly with harmonic order. HHG
	belongs instead to the strong-field, non-perturbative regime: the emitted
	spectrum can contain a broad plateau followed by a cutoff, and the relevant
	observable is the time-dependent microscopic current or polarization induced
	by the field.\cite{ruchon_high_nodate,ghimire_high-harmonic_2019,ortmann_high-harmonic_2021}
	In gases, this phenomenon is often summarized by the three-step picture:
	ionization, acceleration of the liberated electron by the laser field, and
	radiative recombination. This mechanism is central to the generation of XUV
	and attosecond pulses\cite{alexander2025attosecond}, which are used to probe electronic motion on its
	natural time scale.\cite{ruchon_high_nodate}
	
	The observation of HHG in bulk crystals and subsequent work on solid-state
	HHG established crystalline materials as strong-field emitters and as probes
	of ultrafast electronic dynamics in condensed matter.\cite{ghimire_observation_2011,ghimire_high-harmonic_2019}
	In solids, the gas-phase recollision intuition must be modified because the
	electrons occupy Bloch bands rather than an isolated atomic continuum. Under
	a spatially uniform laser field, the crystal momentum is shifted according to
	the acceleration theorem, equivalently through the Peierls substitution
	\(\mathbf{k}\rightarrow\mathbf{k}+e\vec{A}(t)/\hbar\). The resulting HHG signal is
	usually discussed in terms of two coupled mechanisms. The intraband
	mechanism comes from the nonlinear band velocity and the anharmonic current
	generated as carriers move through a non-parabolic band; in strong fields
	this is closely related to ac Bloch oscillations. The interband mechanism
	comes from field-induced electron-hole pairs whose time-dependent
	polarization radiates when the electron and hole recombine. In realistic
	strong-field calculations these contributions are not independent, because
	the same laser field controls both interband population transfer and
	intraband acceleration.\cite{wang2024high,PhysRevB.96.035112}
	
	This sensitivity to band dispersion, gaps, transition matrix elements, and
	crystal symmetry makes solid-state HHG useful beyond light generation. The
	harmonic spectrum can encode information about band structure, driven
	electron-hole motion, Berry curvature and other geometric quantities,
	crystal symmetry, and even real-space lattice or charge-density structure.
	It is also a route toward compact coherent short-wavelength and attosecond
	sources based on solids and engineered nanostructures.\cite{ghimire_high-harmonic_2019,ortmann_high-harmonic_2021}
	For the present work, this motivates a formulation in which the harmonic
	current is obtained directly from a lattice Hamiltonian while keeping the
	sublattice, spin, orbital, and band labels explicit.
	
	The time-dependent Schr\"odinger equation can be solved directly at every crystal momentum and time step, as is done in many HHG calculations \cite{yue_introduction_2022}. The resulting current is then Fourier transformed to obtain the spectrum. 
	For a strictly periodic
	monochromatic drive, however, Floquet theory gives a complementary
	frequency-domain viewpoint. When the Hamiltonian satisfies
	\begin{equation}
		H(t+T,\mathbf{k})=H(t,\mathbf{k}),
		\qquad
		T=\frac{2\pi}{\omega_0},
	\end{equation}
	the solutions can be written as periodic Floquet modes multiplied by
	quasienergy phases\cite{shirley_solution_1965,sambe_steady_1973}.
	Numerically, the Floquet states can be obtained in at least two common
	ways. The first is the stroboscopic construction: one propagates the system
	over one period, diagonalizes the one-period evolution operator \(U(T)\),
	and extracts the quasienergies from its eigenphases. This is close to direct
	time propagation and is useful when the time dependence is most naturally
	handled on a time grid. The second is the Floquet-Sambe construction used
	here: one expands the Hamiltonian and wavefunction in Fourier components
	and converts the periodic Schr\"odinger equation into a static eigenvalue
	problem in an enlarged space of photon replicas. In this representation the
	harmonic structure is explicit, and the Fourier components of the current
	can be related directly to the Floquet eigenvectors.\cite{ikeda_floquet-theoretical_2018,faisal1997floquet}
	
	In this work, we use the Floquet-Sambe representation to derive the current
	harmonics of a periodically driven tight-binding model\cite{jurss2021high,wu2020high}. 
 Within this formulation, HHG can be interpreted in terms of transitions between photon-dressed quasienergy states, offering direct insight into multiphoton processes, selection rules, and plateau formation mechanisms.\\
To the best of our knowledge, there is currently no publicly available and general-purpose code for the scientific community that implements a Floquet-based approach to HHG, particularly within the tight-binding framework and for multi-orbital, multi-dimensional systems. This lack of accessible computational tools has likely contributed to the relatively limited adoption of Floquet methods for HHG, despite their growing importance in related areas.

Indeed, Floquet approaches have recently gained significant traction in ab initio and model-Hamiltonian studies of laser-driven materials, including investigations of multiphoton processes\cite{puente2024computing}, nonlinear optical response functions\cite{alliati2023floquet,10.21468/SciPostPhys.18.6.199}, laser-dressed band structures and Floquet engineering\cite{tiwari2025first},and the interpretation of real-time simulations\cite{giovannini2020floquet}. These developments highlight the timeliness of a flexible and efficient Floquet-based framework tailored specifically to HHG in crystalline systems.\\

Motivated by these considerations, in this work we present a Julia\cite{perkel2019julia} codebase implementing Floquet-Bloch theory for the computation of high-harmonic generation in periodic tight-binding systems. Unlike previous Floquet HHG studies limited to simple models, our implementation is designed to handle complex lattice geometries, multi-orbital unit cells, arbitrary dimensionality (1D, 2D, and 3D), and general light polarizations within the tight-binding approximation. The code provides direct access to quasienergy spectra, time-dependent Floquet states, and the harmonic components of the current, enabling a systematic and transparent analysis of HHG mechanisms in driven quantum materials.
The manuscript is organized as follows. In Sec.~\ref{Theoretical_Methods} we introduce the Floquet formalism starting from a simple one-dimensional tight-binding model and then generalize it to three-dimensional systems and more complex Hamiltonians. In Sec.~\ref{codeval} we validate the implementation by benchmarking against existing results and present illustrative applications to more complex systems in Sec.~\ref{appl}. In Sec.~\ref{usage} we describe the structure of the code and explain its usage, including installation instructions and example workflows. Finally, in Sec.~\ref{conclusions} we summarize our results and outline possible future extensions and applications of the code.









\section{Floquet-Sambe formalism.}
\label{Theoretical_Methods}
	
	We use composite Hamiltonian indices $ \mu \equiv (i,\sigma,\alpha), \nu \equiv (j,\varsigma,\beta). $
	Here \(i,j\) denote site, sublattice, or basis indices,
	\(\sigma,\varsigma\) denote spin indices, and \(\alpha,\beta\) denote orbital
	or band-sector indices. The same Hilbert-space index set is also denoted by
	\(\eta\) and \(\kappa\) when additional dummy indices are required. Matrix
	elements of the Hamiltonian are written as
	\begin{equation}
		H_{\mu\nu}(t,\mathbf{k})
		=
		\mel**{\mu(\mathbf{k})}{H(t,\mathbf{k})}{\nu(\mathbf{k})}.
	\end{equation}
	
	The Sambe space is the extended Hilbert space formed by the tensor product
	of the physical Hilbert space at fixed \(\mathbf{k}\) and the space of
	\(T\)-periodic functions, spanned by the Fourier harmonics
	\(\ee^{\ii n\omega_0 t}\).\cite{sambe_steady_1973} Its basis states are
	therefore labelled by a pair \((n,\mu)\), combining a Fourier index with a
	physical-state index.
	Floquet Fourier indices are denoted by $ n,m,\ell \in [-\NF,\NF]$,
	where \(\NF\) is the numerical truncation of the Sambe space. In the formal
	derivation these sums are infinite; in the implementation they are truncated
	to the range above.
	We begin from the static problem
	\begin{equation}
		H_0(\mathbf{k})\ket{\mu(\mathbf{k})}
		=
		E^0_\mu(\mathbf{k})\ket{\mu(\mathbf{k})}.
		\label{static}
	\end{equation}
	The static eigenvectors form the columns of
	\begin{equation}
		\Phi_0(\mathbf{k})
		=
		\begin{pmatrix}
			\ket{\mu_1(\mathbf{k})}
			&
			\ket{\mu_2(\mathbf{k})}
			&
			\cdots
		\end{pmatrix},
	\end{equation}
	so that
	\begin{equation}
		H_0(\mathbf{k})\Phi_0(\mathbf{k})
		=
		\Phi_0(\mathbf{k})\mathbb{E}_0(\mathbf{k}),
		\qquad
		\left[\mathbb{E}_0(\mathbf{k})\right]_{\mu\nu}
		=
		E^0_\mu(\mathbf{k})\delta_{\mu\nu}.
	\end{equation}
	
	The columns of the exact time-dependent solution matrix \(\Phi(t,\mathbf{k})\)
	satisfy the time dependent Schr\"odinger equation:
	\begin{equation}
		\left[
		H(t,\mathbf{k})-i\hbar\partial_t
		\right]\Phi(t,\mathbf{k})=0.
	\end{equation}
	In order to solve this equation, we use the Floquet ansatz:

	\begin{equation}
			\Phi(t,\mathbf{k})
			=
			\chi(t,\mathbf{k})
			\ee^{-\ii\Xi(\mathbf{k})t/\hbar}
			F(\mathbf{k}).
		\label{eq:floquet_ansatz}
	\end{equation}
	Here \(\chi(t,\mathbf{k})\) is periodic in time,
	\(\chi(t+T,\mathbf{k})=\chi(t,\mathbf{k})\), and
		$ \Xi_{\mu\nu}(\mathbf{k})
		=
		\xi_\mu(\mathbf{k})\delta_{\mu\nu} $
	is the diagonal quasienergy matrix. The constant matrix \(F(\mathbf{k})\) fixes
	the initial condition. Substituting Eq.~\eqref{eq:floquet_ansatz} into the
	Schr\"odinger equation gives the Floquet eigenproblem
	\begin{equation}
			\left[
			H(t,\mathbf{k})-i\hbar\partial_t
			\right]\chi(t,\mathbf{k})
			=
			\chi(t,\mathbf{k})\Xi(\mathbf{k}).
		\label{eq:floquet_operator_problem}
	\end{equation}
	
	The periodic part is expanded on the static eigenbasis as
	\begin{equation}
			\chi(t,\mathbf{k})
			=
			\sum_{n=-\NF}^{\NF}
			\ee^{\ii n\omega_0 t}
			\Phi_0(\mathbf{k})\Lambda^{(n)}(\mathbf{k}).
		\label{eq:chi_expansion}
	\end{equation}
	The matrix \(\Lambda^{(n)}(\mathbf{k})\) has elements
	\begin{equation}
		\left[\Lambda^{(n)}(\mathbf{k})\right]_{\mu\nu}
		=
		\lambda^{(n)}_{\mu\nu}(\mathbf{k}).
	\end{equation}
	The row index of a Sambe-space eigenvector is therefore the pair
	\((n,\mu)\), while the column index \(\nu\) labels the Floquet eigenstate.
	
	At \(t=0\), Eq.~\eqref{eq:floquet_ansatz} gives
	\begin{equation}
		\Phi(0,\mathbf{k})
		=
		\Phi_0(\mathbf{k})
		\left[
		\sum_{n=-\NF}^{\NF}
		\Lambda^{(n)}(\mathbf{k})
		\right]
		F(\mathbf{k}).
	\end{equation}
	If the initial Floquet solution is matched to the static basis,
	\(\Phi(0,\mathbf{k})=\Phi_0(\mathbf{k})\), then
	\begin{equation}
		F(\mathbf{k})
		=
		\Gamma_{\NF}^{-1}(\mathbf{k}),
		\qquad
		\Gamma_{\NF}(\mathbf{k})
		=
		\sum_{n=-\NF}^{\NF}
		\Lambda^{(n)}(\mathbf{k}).
		\label{eq:initial_condition_matrix}
	\end{equation}
	The replacement \(F=\Gamma_{\NF}^\dagger\) is valid as
	\(\Gamma_{\NF}\) is unitary, in the limit of complete basis set. 
	
	The Hamiltonian Fourier series is
	\begin{equation}
		H(t,\mathbf{k})
		=
		\sum_{\ell=-\infty}^{\infty}
		\ee^{\ii\ell\omega_0 t}
		H^{(\ell)}(\mathbf{k}),
	\end{equation}
	with
	\begin{equation}
			H^{(\ell)}_{\mu\nu}(\mathbf{k})
			=
			\frac{1}{T}
			\int_0^T
			dt\,
			\ee^{-\ii\ell\omega_0 t}
			\mel{\mu(\mathbf{k})}{H(t,\mathbf{k})}{\nu(\mathbf{k})}.
		\label{eq:H_fourier_component}
	\end{equation}
	In the numerical implementation, the time integral in
	Eq.~\eqref{eq:H_fourier_component} is evaluated using FFTW, yielding the
	discrete Fourier components in Eq.~\eqref{eq:fft_hamiltonian_component}. The
	procedure is described in detail in Appendix~\ref{app:fft}.
	In the following Sambe construction we distinguish three kinds of indices.
	Fourier-mode indices such as \(m,n,\ell,p\) run over
	\([-\NF,\NF]\). Static internal indices such as \(\mu,\nu,\eta,\kappa\) run over
	the \(H\)-dimensional basis at fixed \(\mathbf{k}\). A full Sambe index is the
	composite pair \((m,\mu)\in[-\NF,\NF]\times[1,H]\); we do not introduce a
	separate symbol for this composite index below. After diagonalizing the
	truncated Sambe Hamiltonian, the retained Floquet basis labels are again
	written with Greek indices in \([1,H]\), corresponding to the quasienergy
	branch \(-\hbar\omega_0/2\le\xi<\hbar\omega_0/2\). The context therefore
	distinguishes Greek indices used as static-basis labels from those used as
	Floquet-mode labels.
	Using Eq.~\eqref{eq:chi_expansion} in
	Eq.~\eqref{eq:floquet_operator_problem} and projecting onto the \(m\)-th
	Floquet component gives
	\begin{equation}
			\sum_{n=-\NF}^{\NF}
			\left[
			H^{(m-n)}(\mathbf{k})
			+
			n\hbar\omega_0 I\,\delta_{mn}
			\right]
			\Lambda^{(n)}(\mathbf{k})
			=
			\Lambda^{(m)}(\mathbf{k})\Xi(\mathbf{k}).
		\label{eq:sambe_matrix_problem}
	\end{equation}
	Thus the Sambe-space Floquet Hamiltonian is
	\begin{equation}
			\left[
			H^F(\mathbf{k})
			\right]_{m\mu,n\nu}
			=
			H^{(m-n)}_{\mu\nu}(\mathbf{k})
			+
			n\hbar\omega_0
			\delta_{mn}\delta_{\mu\nu}.
		\label{eq:sambe_hamiltonian}
	\end{equation}
	The corresponding component eigenvalue equation is
	\begin{equation}
			\sum_{n=-\NF}^{\NF}
			\sum_{\eta}
			\left[
			H^F(\mathbf{k})
			\right]_{m\mu,n\eta}
			\lambda^{(n)}_{\eta\nu}(\mathbf{k})
			=
			\xi_\nu(\mathbf{k})
			\lambda^{(m)}_{\mu\nu}(\mathbf{k}).
		\label{eq:sambe_component_problem}
	\end{equation}
	From the solution of the above system of equations a generic state can be written as: 
	\begin{equation}
		\ket{\psi(t,\mathbf{k})}
		= \sum_\mu w_\mu(\mathbf{k}) \ket{\mu(\mathbf{k})}=
		\Phi(t,\mathbf{k})\vec{w}(\mathbf{k}),
	\end{equation}
	where \(\vec{w}(\mathbf{k})\) is fixed by the initial state. If the initial state
	is the ground state $\ket{\mathrm{G.S.}(\mathbf{k})}$,
	\begin{equation}
		w_\mu(\mathbf{k})
		=
		\braket{\phi_\mu(0,\mathbf{k})}{\mathrm{G.S.}(\mathbf{k})}.
		\label{eq:weights}
	\end{equation}
	Using Eq.~\eqref{eq:initial_condition_matrix}, we define the \textit{Floquet-projected weights}.
	\begin{equation}
			\vec{\gamma}(\mathbf{k})
			=
			\Gamma_{\NF}^{\dagger}(\mathbf{k})\vec{w}(\mathbf{k}).
		\label{eq:gamma_vector}
	\end{equation}
	The state components in the static eigenbasis are then
	\begin{equation}
			\psi_\mu(t,\mathbf{k})
			=
			\sum_{\nu}
			\gamma_\nu(\mathbf{k})
			\ee^{-\ii\xi_\nu(\mathbf{k})t/\hbar}
			\sum_{n=-\NF}^{\NF}
			\ee^{\ii n\omega_0t}
			\lambda^{(n)}_{\mu\nu}(\mathbf{k}).
		\label{eq:state_components}
	\end{equation}
	Using the wave function obtained from Eq.~\eqref{eq:sambe_component_problem}, with the ground state as the initial condition, we can calculate the expectation value of any time-dependent operator.

	\subsection{Harmonic current.}
	In this section we show how to evaluate the current operator on the Floquet states and then how to extract the high-harmonic coefficients. The current operator measured along the Cartesian direction \(a\in\{x,y\}\) is
	\begin{equation}
		\hat{\mathcal I}_a(t,\mathbf{k})
		=
		-\frac{\partial H(t,\mathbf{k})}{\partial A_a(t)}.
		\label{eq:current_operator}
	\end{equation}
	Computationally, the momentum derivative required to evaluate
	Eq.~\eqref{eq:current_operator} is obtained using the fourth-order central
	finite-difference formula in Eq.~\eqref{eq:fourth_order_current_derivative},
	as detailed in Appendix~\ref{app:fft}.
	Its matrix elements in the static basis are
	\begin{equation}
		\mathcal I_{a,\mu\nu}(t,\mathbf{k})
		=
		\mel{\mu(\mathbf{k})}{\hat{\mathcal I}_a(t,\mathbf{k})}{\nu(\mathbf{k})}.
	\end{equation}
	We use the Fourier convention
	\begin{align}
		\mathcal I_a(t,\mathbf{k})
		&=
		\sum_{N=-\infty}^{\infty}
		\ee^{\ii N\omega_0t}
		\mathcal I_a^{(N)}(\mathbf{k}),
		\nonumber\\
		\mathcal I_a^{(N)}(\mathbf{k})
		&=
		\frac{1}{T}
		\int_0^T
		dt\,
		\ee^{-\ii N\omega_0t}
		\mathcal I_a(t,\mathbf{k}).
		\label{eq:current_fourier}
	\end{align}
	
	Combining Eqs.~\eqref{eq:chi_expansion} and \eqref{eq:current_fourier}
	gives
	\begin{align}
		\chi^\dagger(t,\mathbf{k})
		\mathcal I_a(t,\mathbf{k})
		\chi(t,\mathbf{k})
		&=
		\sum_{m,n,L}
		\ee^{\ii(n+L-m)\omega_0t}
		\Lambda^{(m)\dagger}(\mathbf{k})
		\mathcal I_a^{(L)}(\mathbf{k})
		\Lambda^{(n)}(\mathbf{k})
		\nonumber\\
		&=
		\sum_{N=-\infty}^{\infty}
		\ee^{\ii N\omega_0t}
		\mathcal J_a^{(N)}(\mathbf{k}),
	\end{align}
	where
	\begin{equation}
			\mathcal J_a^{(N)}(\mathbf{k})
			=
			\sum_{m,n=-\NF}^{\NF}
			\Lambda^{(m)\dagger}(\mathbf{k})
			\mathcal I_a^{(N-n+m)}(\mathbf{k})
			\Lambda^{(n)}(\mathbf{k}).
		\label{eq:rotated_current_harmonic}
	\end{equation}
	Terms whose Floquet or current Fourier indices fall outside the numerical
	truncation are set to zero. In components,
	\begin{equation}
			\left[
			\mathcal J_a^{(N)}(\mathbf{k})
			\right]_{\mu\nu}
			=
			\sum_{m,n}
			\sum_{\eta,\kappa}
			\lambda^{(m)*}_{\mu \eta}(\mathbf{k})
			\mathcal I^{(N-n+m)}_{a,\eta\kappa}(\mathbf{k})
			\lambda^{(n)}_{\kappa\nu}(\mathbf{k}).
		\label{eq:rotated_current_component}
	\end{equation}
	
	The total current is
	\begin{equation}
	\label{eq:TC}
		I_a(t)
		=
		\sum_{\mathbf{k}}
		\mel{\psi(t,\mathbf{k})}
		{\hat{\mathcal I}_a(t,\mathbf{k})}
		{\psi(t,\mathbf{k})}.
	\end{equation}
	Substituting the Floquet state yields
	\begin{equation}
				I_a(t)
				=
				\sum_{\mathbf{k}}
				\sum_{N,\mu,\nu}
				\gamma^\star_\mu(\mathbf{k})
				\gamma_\nu(\mathbf{k})
				\left[
				\mathcal J_a^{(N)}(\mathbf{k})
				\right]_{\mu\nu}
				\ee^{
					\ii
					\left[
					N\omega_0
					+
					\frac{\xi_\mu(\mathbf{k})-\xi_\nu(\mathbf{k})}{\hbar}
					\right]t
				}.
		\label{eq:full_current}
	\end{equation}
	The diagonal part \(\mu=\nu\) contains only integer harmonics of the driving
	frequency:
	\begin{equation}
			I^{(H)}_a(t)
			=
			\sum_{\mathbf{k}}
			\sum_{N=-\infty}^{\infty}
			\ee^{\ii N\omega_0t}
			\sum_\mu
			|\gamma_\mu(\mathbf{k})|^2
			\left[
			\mathcal J_a^{(N)}(\mathbf{k})
			\right]_{\mu\mu}.
		\label{eq:harmonic_current}
	\end{equation}
	The off-diagonal part \(\mu\neq \nu\) gives coherent quasienergy beat
	frequencies. The long-time projection of the \(N\)-th harmonic is
	\begin{equation}
		I_{a,N}
		=
		\lim_{T\to\infty}
		\frac{1}{T}
		\int_0^{T}
		dt\,
		\ee^{-\ii N\omega_0t}
		I_a(t).
		\label{eq:time_int_harmonic}
	\end{equation}
	Thus the harmonic contribution is
	\begin{equation}
			I_{a,N}^{(H)}
			=
			\sum_{\mathbf{k}}
			\sum_\mu
			|\gamma_\mu(\mathbf{k})|^2
			\left[
			\mathcal J_a^{(N)}(\mathbf{k})
			\right]_{\mu\mu}
		\label{eq:long_time_harmonic}
	\end{equation}
	Numerically, the time projection in Eq.~\eqref{eq:time_int_harmonic} is
	evaluated using FFTW according to Eq.~\eqref{eq:fft_hhc_spectrum}; further
	details are given in Appendix~\ref{app:fft}. For the harmonic spectra reported
	here, the coherent off-diagonal contribution is omitted by retaining only the
	diagonal Floquet-population contribution in Eq.~\eqref{eq:long_time_harmonic}.
	
	\subsection{Tight-binding implementation.}
	The formalism presented in the previous section is general; we shall now focus on a tight-binding Hamiltonian implementation of the Floquet-Sambe formalism. The static tight-binding Hamiltonian is written as
	\begin{equation}
		H^0_{\mu\nu}(\mathbf{k})
		=
		\sum_{\vec{R}}
		\ee^{\ii\mathbf{k}\cdot\vec{d}_{\vec{R},\mu\nu}}
		\mathbb{H}_{\vec{R},\mu\nu},
		\qquad
		\vec{d}_{\vec{R},\mu\nu}
		=
		\vec{R}+\vec{\delta}_{\mu\nu}.
		\label{eq:static_tb_hamiltonian}
	\end{equation}
	The drive is introduced through the Peierls substitution
	\begin{equation}
		\mathbf{k}
		\rightarrow
		\mathbf{k}
		+
		\frac{q_e}{\hbar}\vec{A}(t),
	\end{equation}
	which gives
	\begin{equation}
		H_{\mu\nu}(t,\mathbf{k})
		=
		\sum_{\vec{R}}
		\ee^{\ii
			\left[
			\mathbf{k}\cdot\vec{d}_{\vec{R},\mu\nu}
			+
			\frac{q_e}{\hbar}\vec{A}(t)\cdot\vec{d}_{\vec{R},\mu\nu}
			\right]}
		\mathbb{H}_{\vec{R},\mu\nu}.
		\label{eq:peierls_hamiltonian}
	\end{equation}
	For simplicity, we restrict the following discussion to monochromatic,
	sinusoidally varying, linearly polarised fields:
	\begin{align}
		\vec{A}_{x}(t)
		=
		-\frac{E_0}{\omega}\sin(\omega t)\hat{x}
		\mbox{ , }
		\vec{A}_{y}(t)
		=
		-\frac{E_0}{\omega}\sin(\omega t)\hat{y}.
		\label{eq:vector_potential_simple}
	\end{align}
	With this choice, \(\vec{A}(0)=0\) and \(H(t=0,\mathbf{k})=H^0(\mathbf{k})\).
	
	For each hopping vector and linear polarisation parameter
	\(P\in\{x,y\}\), we write
	\begin{equation}
		\vec{A}_P(t)
		=
		-\frac{E_0}{\omega}\sin(\omega t)\hat{\epsilon}_P,
		\qquad
		\hat{\epsilon}_1=\hat{x},
		\quad
		\hat{\epsilon}_2=\hat{y}.
	\end{equation}
	The Peierls phase is then
	\begin{equation}
		\frac{q_e}{\hbar}\vec{A}_P(t)\cdot\vec{d}_{\vec{R},\mu\nu}
		=
		-G^{(P)}_{\vec{R},\mu\nu}\sin(\omega t),
		\label{eq:bessel_phase}
	\end{equation}
	where the dimensionless driving amplitude is the projection of the hopping
	vector onto the polarisation direction,
	\begin{equation}
		G^{(P)}_{\vec{R},\mu\nu}
		=
		\frac{q_eE_0}{\hbar\omega}
		\hat{\epsilon}_P\cdot\vec{d}_{\vec{R},\mu\nu}
		=
		\frac{q_eE_0}{\hbar\omega}
		\sum_{b=x,y}\epsilon_P^b d^b_{\vec{R},\mu\nu}.
	\end{equation}
	For the two linear polarisations used here this reduces to the corresponding
	Cartesian component of the hopping vector. By inserting the modified Peierls
	Hamiltonian of Eq.~\eqref{eq:peierls_hamiltonian} into
	Eq.~\eqref{eq:H_fourier_component} and using the Jacobi-Anger expansion, we obtain
	\begin{equation}
			H^{(\ell,P)}_{\mu\nu}(\mathbf{k})
			=
			\sum_{\vec{R}}
			\ee^{\ii\mathbf{k}\cdot\vec{d}_{\vec{R},\mu\nu}}
			(-1)^\ell
			J_\ell\!\left(G^{(P)}_{\vec{R},\mu\nu}\right)
			\mathbb{H}_{\vec{R},\mu\nu}.
		\label{eq:bessel_hamiltonian}
	\end{equation}
	The factor \((-1)^\ell\) follows from
	\(J_\ell(-G)=(-1)^\ell J_\ell(G)\).
	In the static eigenbasis, the Sambe block used in Eq.~\eqref{eq:sambe_hamiltonian}
	is
	\begin{equation}
		H^F_{mn}(\mathbf{k})
		=
		\Phi^\dagger_0(\mathbf{k})
		H^{(m-n,P)}(\mathbf{k})
		\Phi_0(\mathbf{k})
		+
		n\hbar\omega\delta_{mn}\mathbb{I}.
		\label{eq:sambe_static_basis}
	\end{equation}
	
	The same expansion gives the Fourier components of the current measured
	along a Cartesian direction \(a\in\{x,y\}\). From
	Eq.~\eqref{eq:current_operator},
	\begin{equation}
		\hat{\mathcal I}_a(t,\mathbf{k})
		=
		-\ii\frac{q_e}{\hbar}
		\sum_{\vec{R}}
		\ee^{\ii
			\left[
			\mathbf{k}\cdot\vec{d}_{\vec{R},\mu\nu}
			+
			\frac{q_e}{\hbar}\vec{A}_P(t)\cdot\vec{d}_{\vec{R},\mu\nu}
			\right]}
		\mathbb{H}_{\vec{R},\mu\nu}
		d^a_{\vec{R},\mu\nu}.
	\end{equation}
	Therefore the \(N\)-th Fourier component of the current operator is
	\begin{equation}
			\mathcal I^{(N,P)}_{a,\mu\nu}(\mathbf{k})
			=
			-\ii\frac{q_e}{\hbar}
			\sum_{\vec{R}}
			\ee^{\ii\mathbf{k}\cdot\vec{d}_{\vec{R},\mu\nu}}
			(-1)^N
			J_N\!\left(G^{(P)}_{\vec{R},\mu\nu}\right)
			\mathbb{H}_{\vec{R},\mu\nu}
			d^a_{\vec{R},\mu\nu}.
		\label{eq:bessel_current}
	\end{equation}
	Here \(N\) labels the current Fourier component. In the Floquet-mode
	convolution of Eq.~\eqref{eq:rotated_current_harmonic}, the current component
	entering the emitted \(N\)-th harmonic has index \(N-n+m\).
	
	Equations~\eqref{eq:rotated_current_harmonic} and
	\eqref{eq:long_time_harmonic} give the high-harmonic current amplitude
	represented in the spectra discussed in the next section.

	\section{Code validation}
\label{codeval}
This section benchmarks our implementation of the Floquet formalism and harmonic current extraction derived in Sec.~\ref{Theoretical_Methods}. 
In particular, we validate the numerical construction and truncation of the Floquet Hamiltonian, the reconstruction of time-periodic Floquet states, and the evaluation of the harmonic current \(I_H(N)\) from the microscopic current operator. 
The validation is performed against the canonical one-dimensional two-band model studied by Ref.~\cite{ikeda_floquet-theoretical_2018}, which provides both analytical limits and representative numerical spectra across weak- and strong-field regimes.
	The Bloch Hamiltonian used by Ikeda \textit{et al.}~\cite{ikeda_floquet-theoretical_2018} is written as
	\begin{equation}
		H_0(k)
		=
		2t_0\cos k\,\sigma_x
		+
		Q\sigma_z.
		\label{eq:ikeda_chain_bloch}
	\end{equation}
	with \(t_0=1/2\), we have \(H_0(k)=\cos k\,\sigma_x+Q\sigma_z\),
	with a direct gap \(2|Q|\) at the Brillouin-zone boundary. For a
	monochromatic electric field \(E(t)=-E_0\cos(\omega t)\), the vector
	potential is \(A(t)=A_0\sin(\omega t)\), and the Peierls substitution gives
	\begin{equation}
		H(k,t)
		=
		2t_0\cos\!\left[k+F\sin(\omega t)\right]\sigma_x
		+
		Q\sigma_z,
		\qquad
		F=\frac{q_eaE_0}{\hbar\omega}
		=
		\frac{\Omega_{\mathrm B}}{\omega}.
		\label{eq:ikeda_chain_driven}
	\end{equation}
	This form is useful for checking the present Fourier--Floquet
	implementation because every \(k\) remains independent and the Fourier
	components of the Peierls phase are controlled by Bessel functions.
	
\begin{figure}[H]
		\centering
		\begin{subfigure}[t]{0.48\linewidth}
			\centering
			\includegraphics[width=\linewidth]{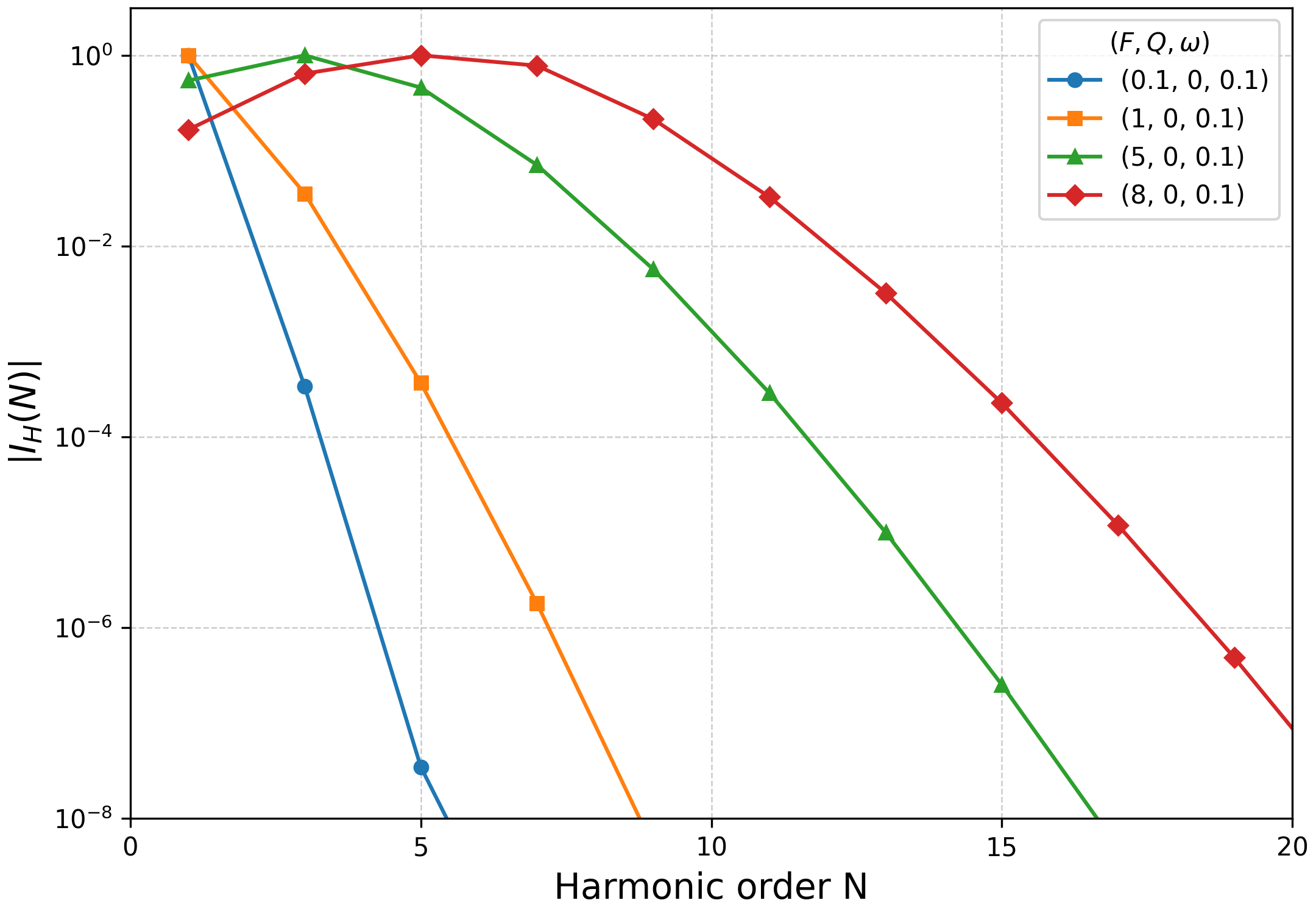}
			\caption{Present calculation: \(Q=0\).}
			\label{fig:nn-q0-omega-01}
		\end{subfigure}\hfill
		\begin{subfigure}[t]{0.48\linewidth}
			\centering
			\includegraphics[width=\linewidth]{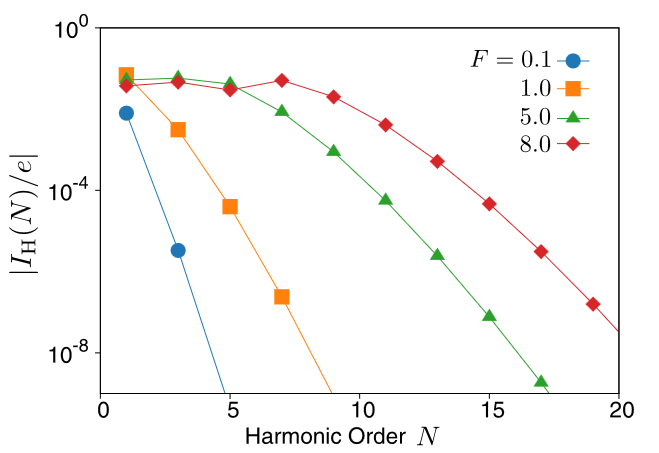}
			\caption{Ikeda \emph{et al.}, Fig.~2: \(Q=0\).}
			\label{fig:nn-q0-reference}
		\end{subfigure}
		\par\medskip
		\begin{subfigure}[t]{0.48\linewidth}
			\centering
			\includegraphics[width=\linewidth]{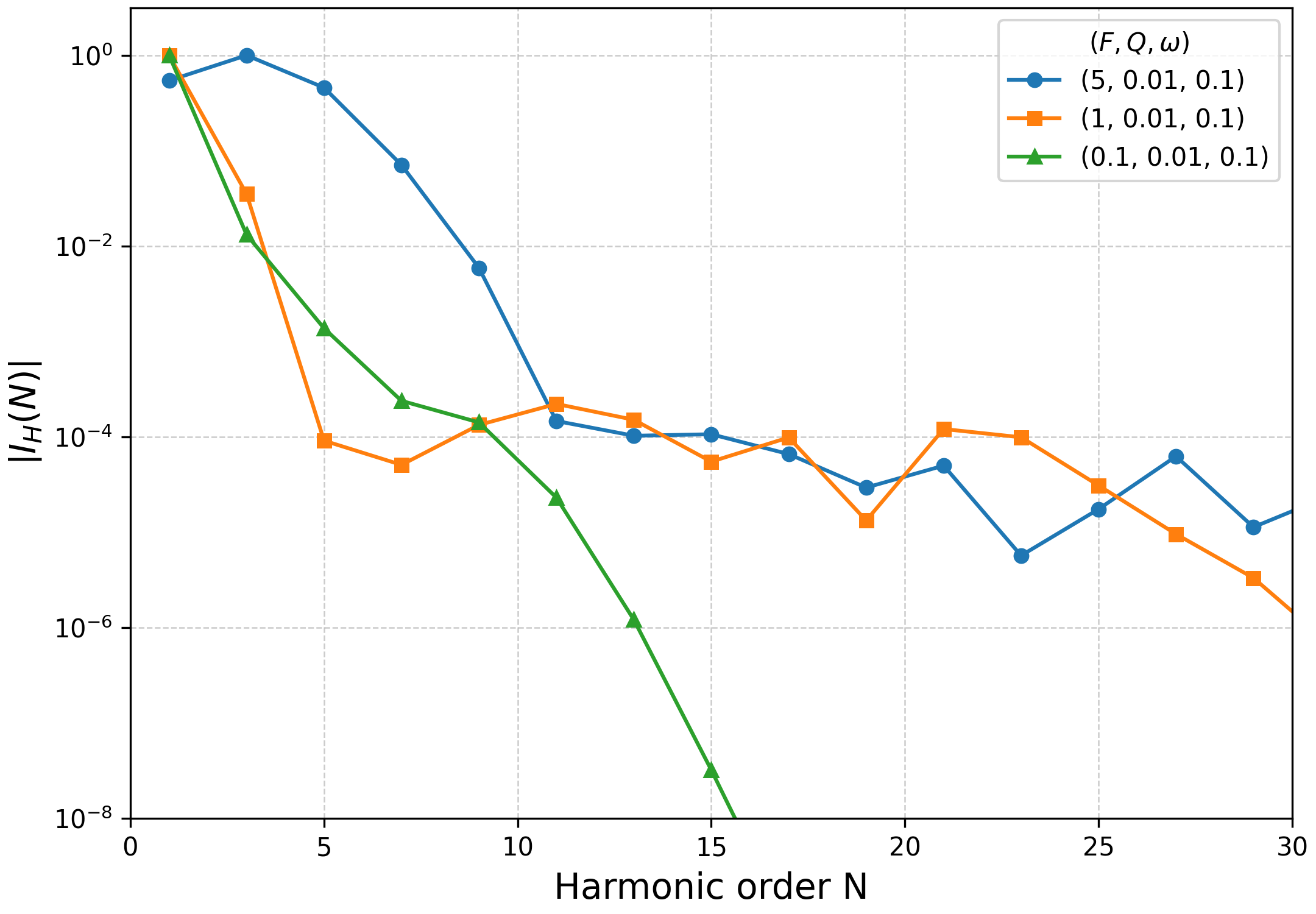}
			\caption{Present calculation: \(Q=0.01\).}
			\label{fig:nn-nonperturbative-q-calculated}
		\end{subfigure}\hfill
		\begin{subfigure}[t]{0.48\linewidth}
			\centering
			\includegraphics[width=\linewidth]{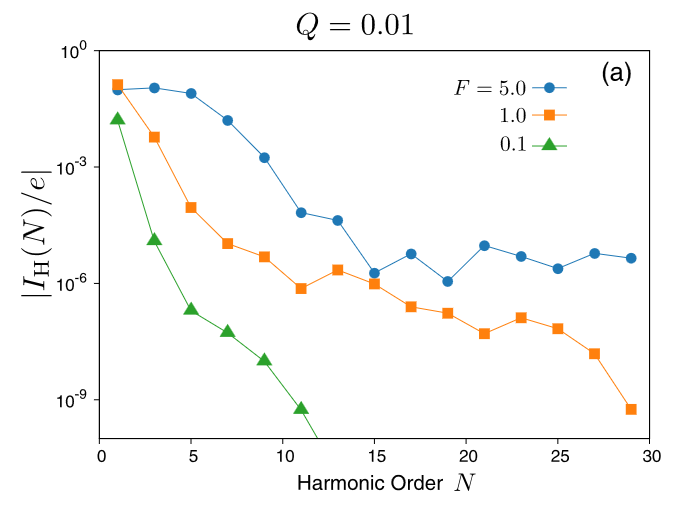}
			\caption{Ikeda \emph{et al.}, Fig.~5(a): \(Q=0.01\).}
			\label{fig:nn-nonperturbative-q-reference}
		\end{subfigure}
		\caption{Comparison of the present HHC spectra (left column) with the
		corresponding reference spectra of Ikeda \emph{et al.}~\cite{ikeda_floquet-theoretical_2018}
		(right column), for \(\omega=0.1\). The upper row shows \(Q=0\) with
		\(F=0.1,1,5,8\); the lower row shows \(Q=0.01\) with \(F=0.1,1,5\).
		Reference panels are reproduced from their Figs.~2 and 5(a), respectively.
		The original ordinate labels and scales are retained; the panels should
		therefore be compared with these differences in mind.}
		\label{fig:ikeda-benchmark-comparison}
	\end{figure}

	The observations from Ref.~\cite{ikeda_floquet-theoretical_2018} that are
	most relevant for the spectra below are the following. First, the measured
	long-time high-harmonic component is not determined by the Floquet bands
	alone. The overlap of the initial half-filled ground state with each basis
	state defines the weights \(w_\mu(\mathbf{k})\) in Eq.~\eqref{eq:weights}.
	These weights are projected onto the Floquet eigenbasis through
	Eq.~\eqref{eq:gamma_vector}, yielding the amplitudes
	\(\gamma_\mu(\mathbf{k})\). Their squared moduli
	\(\lvert\gamma_\mu(\mathbf{k})\rvert^2\) are the Floquet-state population
	factors retained in Eq.~\eqref{eq:long_time_harmonic}. Second, the symmetries
	of the Hamiltonian in Eq.~\eqref{eq:bessel_hamiltonian} lead to HHC spectra
	dominated by odd-order harmonics. Third, the \(Q=0\) single-band
	limit already produces a plateau because the lattice current is nonlinear
	in the Peierls phase; its width is set by the dimensionless field amplitude
	\(|F|\), Eq.~\eqref{eq:ikeda_chain_driven}. To benchmark our code,
	Fig.~\ref{fig:ikeda-benchmark-comparison} places the present calculations
	alongside the corresponding spectra of Ikeda \emph{et al.}~\cite{ikeda_floquet-theoretical_2018}.
	The upper row compares the \(Q=0\) case, with a single plateau, and the
	lower row compares the \(Q=0.01\) case, where additional plateau structures appear.
	
All calculations have been performed using $101$~k-points and up to $80$ Floquet modes. To verify convergence, we also compared the Floquet eigenstates and weights with those reported by Ikeda et al., \cite{ikeda_floquet-theoretical_2018} finding a very close agreement. 

	\section{Applications} \label{appl}
	In this section we present two applications of our code to two different Hamiltonians: a tight binding Hamiltonian for gapped graphene and the 2D Kane-Mele model. Thanks to the efficiency of the Floquet formulation and its fast implementation in Julia, our code allows us to map the HHG response as a function of the model parameters as shown in following sections.

	\subsection{Gapped graphene}
\begin{figure}[H]
    \centering
    \begin{subfigure}[b]{0.48\linewidth}
        \centering
        \includegraphics[width=\linewidth]{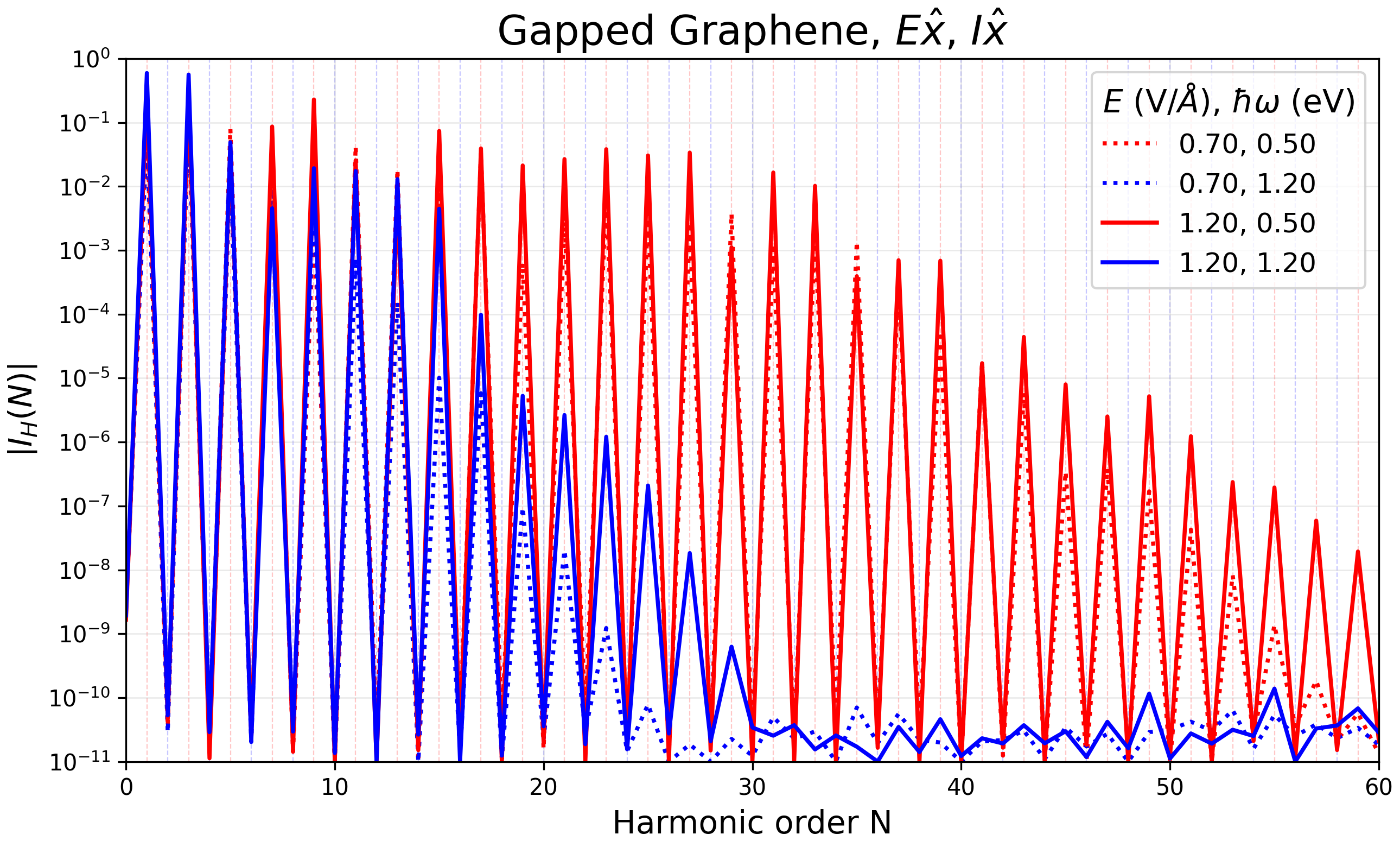}
        \caption{Polarization along the $x$-direction.}
        \label{fig:hBN-sub1}
    \end{subfigure}
    \hfill 
    \begin{subfigure}[b]{0.48\linewidth}
        \centering
        \includegraphics[width=\linewidth]{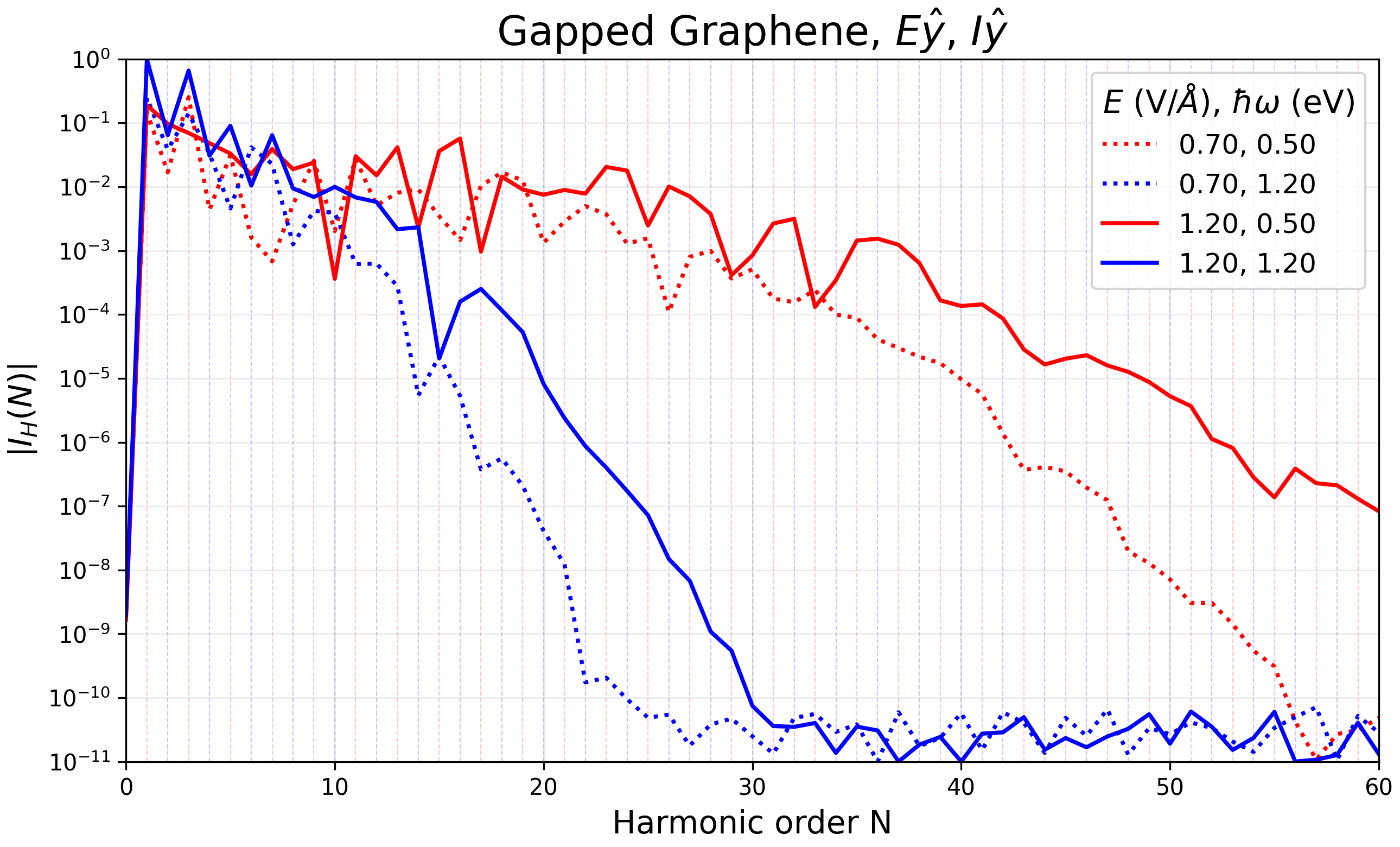} 
        \caption{Polarization along the $y$-direction.}
        \label{fig:hBN-sub2}
    \end{subfigure}

    \caption{High-harmonic spectra of gapped graphene, shown as a function of electric-field amplitude and laser frequency. The field polarization and measured current are shown for both (a) $x$-direction and (b) $y$-direction.}
    \label{fig:hBN-Q-HHC}
\end{figure}

	For gapped graphene we consider a tight binding model with nearest-neighbour hopping in the form
		\begin{equation}
			H_{0}(\mathbf{k}) = -t_0 \sum_{\expval{i,j}} \left(c^\dagger_i c_j + \text{h.c.}\right) + Q \sum_j (-1)^j c^\dagger_j c_j \;,\; Q = \frac{E_g}{2}
		\end{equation}
		In Fig.~\ref{fig:hBN-Q-HHC} we show the HHG response of gapped graphene along both the $x$ and $y$  directions for an external field in the same directions at different frequencies and intensities. In this example, we used tight binding parameters $E_g = 3.92~\mathrm{eV}$ and $ t_0 = 2.42~\mathrm{eV}$.
As shown in Fig.~\ref{fig:hBN-Q-HHC}(a), only odd harmonics are present in the $x$ direction due to the $x \rightarrow -x$ symmetry. Conversely, as seen in panel (b) of the same figure, all harmonics are allowed by symmetry along the $y$ direction.

	\begin{figure}[H]
			\centering
			\includegraphics[width=0.95\linewidth]{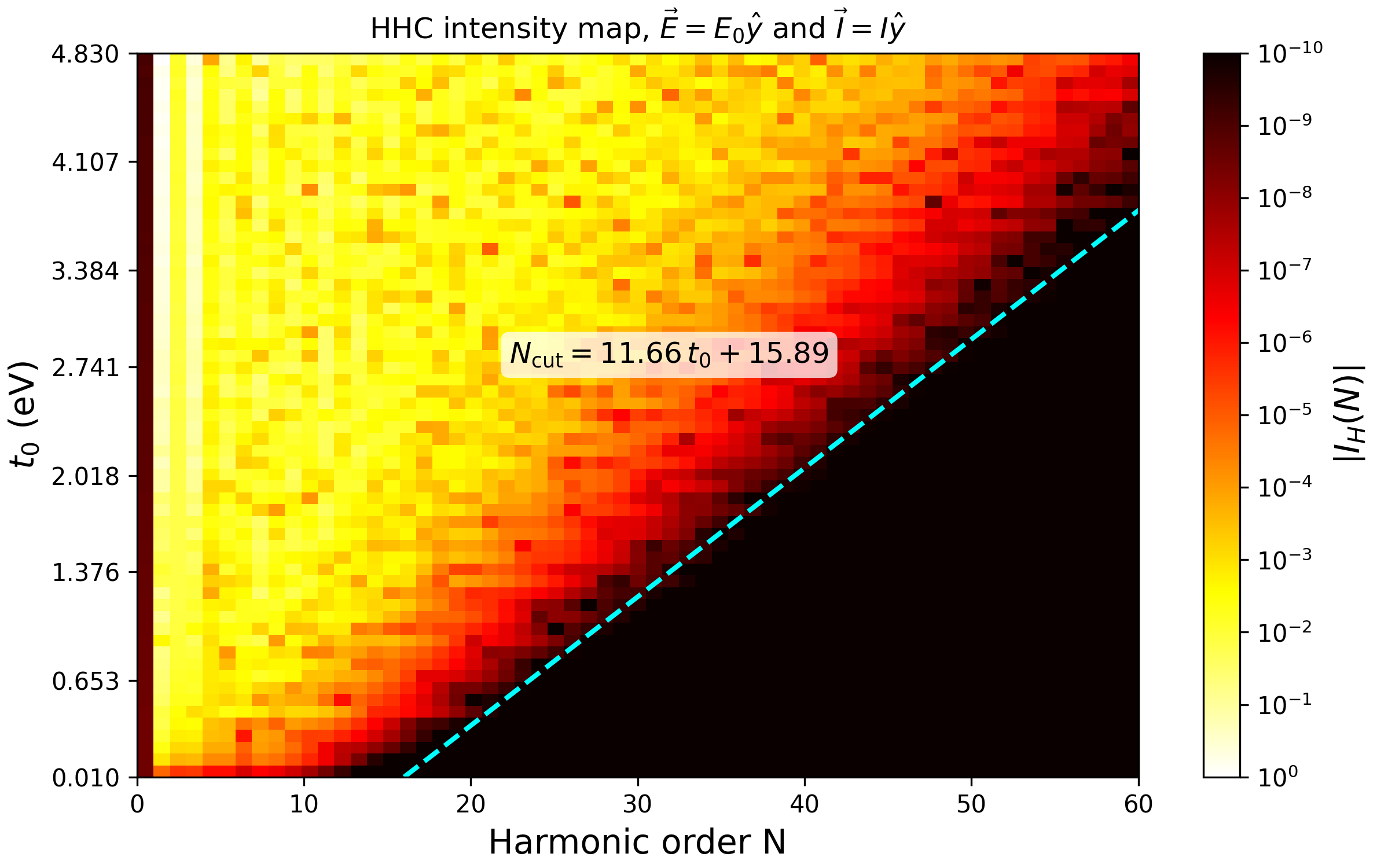}
			\par\smallskip
			(b) \(y\)-polarized field.
		\caption{Longitudinal high-harmonic current spectrum of gapped graphene as a function of \(t_0\) (\(\mathrm{eV}\))}
		\label{fig:hbn-t0-longitudinal}
	\end{figure}

The HHC phase diagrams also summarize the symmetry selection rules of the
response. In the longitudinal geometry, where the driving field and measured
current are parallel, an \(x\)-polarized field measured through \(I_x\) produces
only odd harmonics. For a \(y\)-polarized field measured through \(I_y\), the
gapped honeycomb model allows mixed odd and even harmonics, whereas gapless
graphene retains an odd-harmonic response. In the transverse geometry, the
\((E_x,I_y)\) channel shows even harmonics in the gapped model but is strongly
suppressed in graphene, while the \((E_y,I_x)\) channel is negligible in both
cases on the plotted scale.

Figure~\ref{fig:hbn-t0-longitudinal} illustrates how the Floquet approach can
be used to map the HHG response as a function of microscopic parameters. Here
the onsite term, electric-field amplitude, and photon energy are kept fixed,
while the hopping parameter \(t_0\) is varied. The intensity is plotted on a
logarithmic scale for each value of \(t_0\). Increasing \(t_0\) broadens the
high-harmonic plateau and shifts the cutoff to higher harmonic order. This
trend is consistent with the cutoff estimate obtained in
Ref.~\cite{ikeda_floquet-theoretical_2018} that found the relation:
\begin{equation}
	N_{\mathrm{cut}}
	=
	\frac{8}{\pi}\abs{J_1\left(\frac{q_eaE_0}{\hbar \omega}\right)}\frac{t_0}{\omega},
\end{equation}
where \(N_{\mathrm{cut}}\) is the harmonic cutoff order and the corresponding
cutoff energy is \(E_{\mathrm{cut}} = N_{\mathrm{cut}}\hbar\omega.\) \\
The boundary of the bright plateau in Fig.~\ref{fig:hbn-t0-longitudinal}
therefore follows the expected approximately linear scaling with \(t_0\) at
fixed \(E_0\) and \(\omega\). Following the same procedure, the HHG response can
be systematically investigated as a function of other key parameters, such as
the energy gap, field intensity, and laser frequency found a perfect agreement with the prediction of Ikeda et al.\cite{ikeda_floquet-theoretical_2018} .

		\subsection{Kane-Mele Model}
		The second example is the 2D Kane-Mele model, whose Hamiltonian is
		decomposed into four contributions:
		\begin{equation}
			H_{0} = H_{\text{hopping}} + H_{\text{SOC}} + H_{\text{Rashba}} + H_{\text{Onsite}} 
		\end{equation}
		The nearest-neighbour hopping contribution is
		\begin{equation}
			H_{\text{hopping}}
			= -t_0\sum_{\langle i,j\rangle,\sigma}
			\left(c^\dagger_{i\sigma}c_{j\sigma}+\text{h.c.}\right).
		\end{equation}
		Here \(t_0\) is the hopping amplitude, \(\langle i,j\rangle\) denotes
		nearest-neighbour pairs, and \(c^\dagger_{i\sigma}\)
		(\(c_{i\sigma}\)) creates (annihilates) an electron of spin \(\sigma\)
		at site \(i\). The abbreviation h.c. denotes the Hermitian conjugate.

		The intrinsic spin-orbit coupling contribution is
		\begin{equation}
			H_{\text{SOC}}
			= i\lambda\sum_{\langle\langle i,j\rangle\rangle,\sigma,\varsigma}
			\nu_{ij}c^\dagger_{i\sigma}(s_z)_{\sigma\varsigma}c_{j\varsigma}.
		\end{equation}
		Here \(\lambda\) is the intrinsic spin-orbit coupling strength,
		\(\langle\langle i,j\rangle\rangle\) denotes next-nearest-neighbour pairs,
		and \(s_z\) is the Pauli matrix acting on spin. The chirality factor
		\(\nu_{ij}=\pm1\) changes sign between clockwise and counter-clockwise
		hopping paths. Because this contribution is proportional to \(s_z\),
		the two spin sectors acquire opposite Haldane masses.

		The Rashba spin-orbit coupling contribution is
		\begin{equation}
			H_{\text{Rashba}}
			= \gamma\sum_{\langle i,j\rangle,\sigma,\varsigma}
			c^\dagger_{i\sigma}
			\left([\vec{s}\times\hat{\delta}_j]_z\right)_{\sigma\varsigma}
			c_{j\varsigma}.
		\end{equation}
		Here \(\gamma\) is the Rashba coupling strength,
		\(\vec{s}=(s_x,s_y,s_z)\) is the vector of spin Pauli matrices, and
		\(\hat{\delta}_j\) denotes the unit vector along the corresponding
		nearest-neighbour bond.

		Finally, the staggered onsite contribution is
		\begin{equation}
			H_{\text{Onsite}}
			= Q\sum_{j,\sigma}(-1)^j c^\dagger_{j\sigma}c_{j\sigma}.
		\end{equation}
		The factor \((-1)^j\) assigns opposite signs to the two sublattices,
		giving onsite energies \(+Q\) and \(-Q\). In the absence of spin-orbit
		coupling, the corresponding band gap is \(E_g=2Q\).
		The parameters \(t_0\), \(\lambda\), \(\gamma\), and \(Q\) are expressed in eV.

\begin{figure}[H]
			\centering
			\begin{minipage}{0.85\linewidth}
				\centering
				\includegraphics[width=\linewidth,height=0.25\textheight,keepaspectratio]{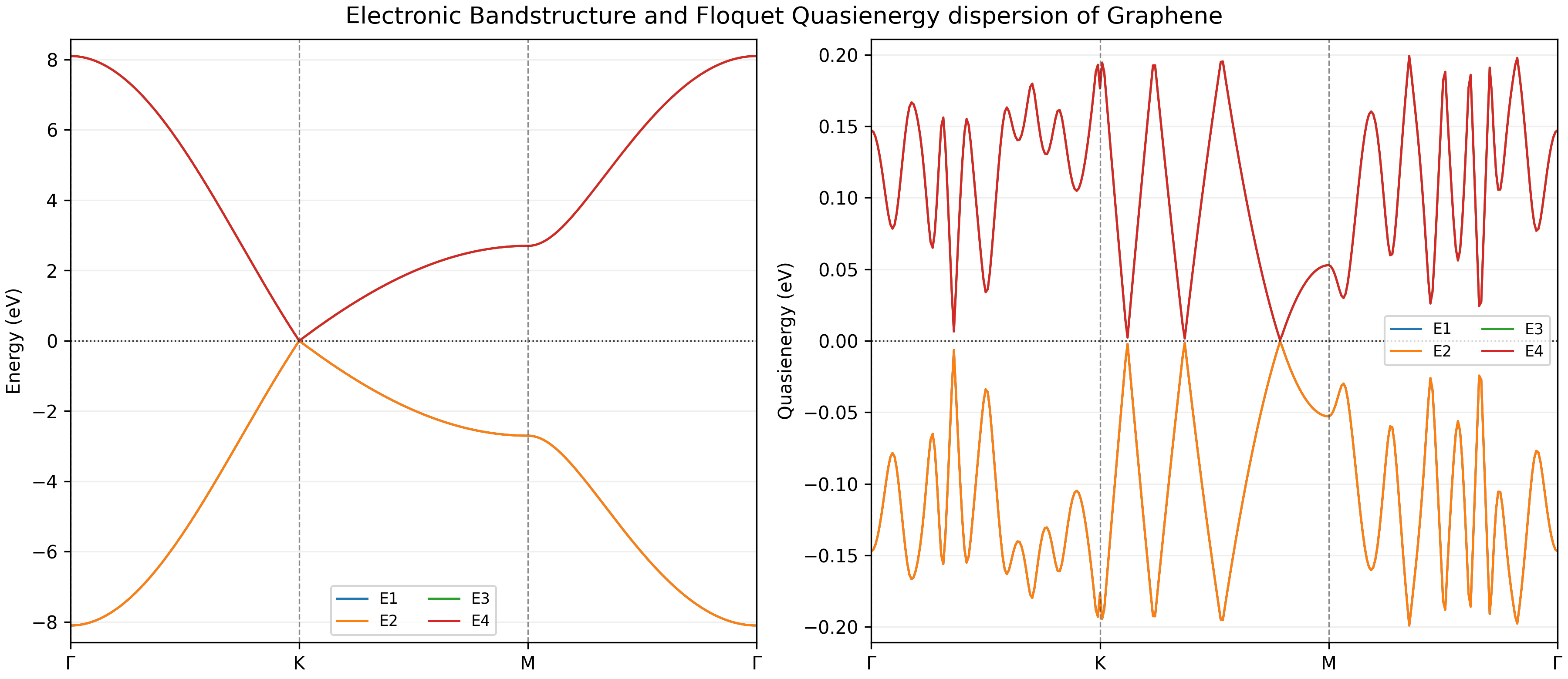}
				\par\vspace{2pt}
				\footnotesize (a) Graphene with \(E_g=0\), \(\lambda=0\).
			\end{minipage}
			\par\vspace{1em}
			\begin{minipage}{0.85\linewidth}
				\centering
				\includegraphics[width=\linewidth,height=0.25\textheight,keepaspectratio]{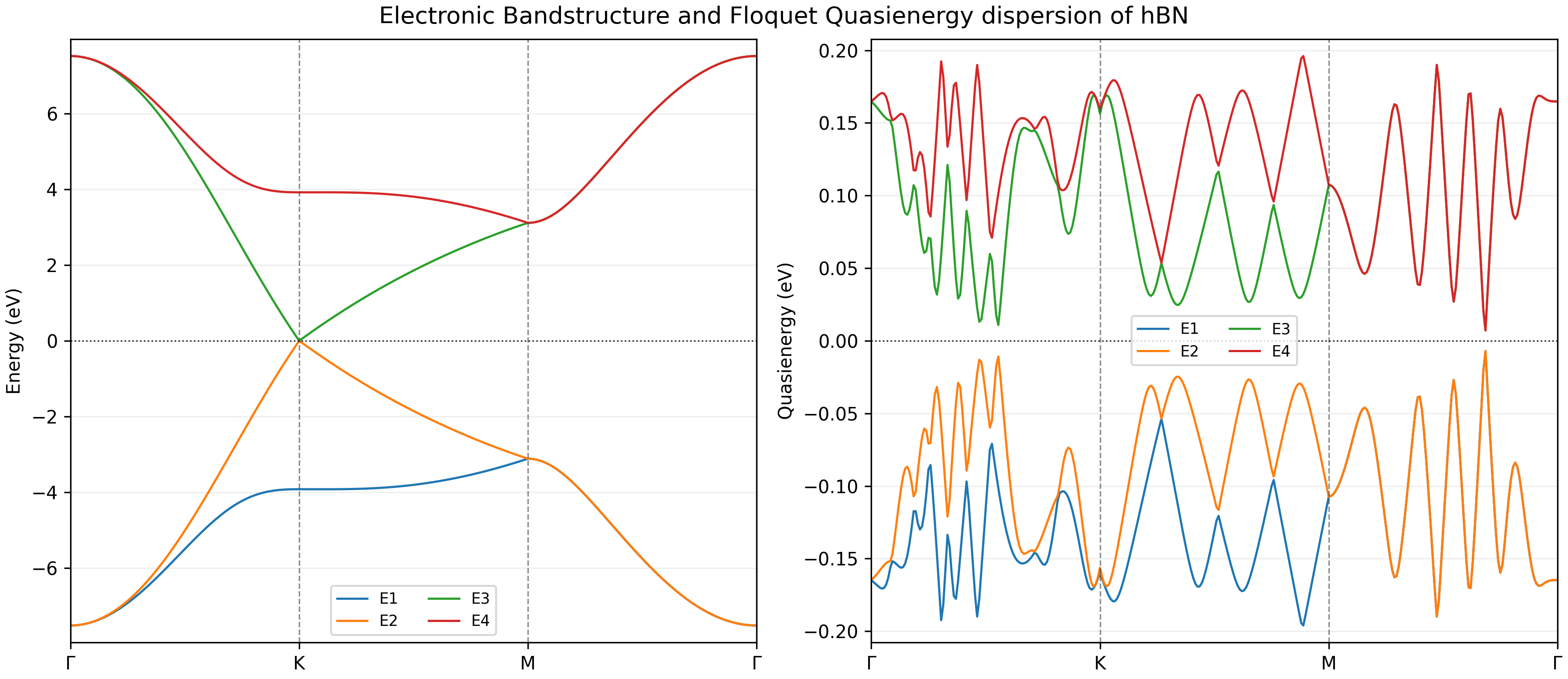}
				\par\vspace{2pt}
				\footnotesize (b) Gap-reclosed hBN with \(E_g=3.92~\mathrm{eV}\), \(\lambda=\lambda^\star\).
			\end{minipage}
		\caption{Static reference band structures for graphene and the gap-reclosed hBN model.}
		\label{fig:haldane_static_band_structures}
	\end{figure}

		We compare two gapless static reference systems. For graphene, we select
		\(E_g=0\), \(t_0=2.7~\mathrm{eV}\), and \(\lambda=0\).
		For the hBN-based model, we use \(E_g=3.92~\mathrm{eV}\),
		\(t_0=2.42~\mathrm{eV}\), and
		\(\lambda=\lambda^\star=E_g/(6\sqrt{3})=0.1556\,t_0\).
		The value \(\lambda^\star\) is chosen to close the static gap:
		$ E_{\mathrm{gap}}^{0}(\lambda^\star) = \min_{\mathbf{k}} \left[ E_c^0(\mathbf{k};\lambda^\star) - E_v^0(\mathbf{k};\lambda^\star) \right] = 0. $\\
	Thus graphene is gapless by honeycomb symmetry in the nearest-neighbour
	limit, whereas the hBN is made gapless by compensating the
	staggered sublattice mass with the complex next-nearest-neighbour term.
	First of all we now compare field-induced quasienergy gaps in the driven Haldane-type
	model. The static reference band structures are shown in
	Fig.~\ref{fig:haldane_static_band_structures}. Graphene is gapless in the
	nearest-neighbour limit, while the hBN-based model is made gapless by the
	critical coupling \(\lambda^\star\). Therefore, the gaps reported below are
	induced by the periodic drive rather than inherited from the static
	Hamiltonian.
	
	The field-induced quasienergy gaps are summarized in
	Table~\ref{tab:floquet_gap_linear_xy}. For the parameters shown here,
	graphene develops larger Floquet gaps than the gap-reclosed hBN model for
	both linear polarizations.
	
	\begin{table}[htbp]
		\centering
		\caption{Largest observed Floquet quasienergy gaps among the calculated parameter values for linearly polarized electric fields along \(x\) and \(y\). The values of \(E_0\) and \(\omega\) specify the field amplitude and photon energy at which each maximum gap is obtained.}
		\label{tab:floquet_gap_linear_xy}
		\begin{tabular}{llccc}
			\toprule
			Polarization & System & \(E_0\) (V/nm) & \(\omega\) (eV) & Floquet gap (meV) \\
			\midrule
			Linear-\(y\) & Graphene (c6) & \(11.79\) & \(2.03\) & \(40.28\) \\
			Linear-\(y\) & hBN           & \(5.07\) & \(1.00\) & \(5.93\)  \\
			\midrule
			Linear-\(x\) & Graphene (c6) & \(10.97\) & \(2.03\) & \(34.28\) \\
			Linear-\(x\) & hBN           & \(7.00\) & \(0.22\) & \(5.23\)  \\
			\bottomrule
		\end{tabular}
	\end{table}
	
	This shows that two systems with zero static gap can nevertheless have quantitatively different driven
	quasienergy spectra.
	However, tuning the hBN-based model to \(\lambda^\star\) does not make its
	Hamiltonian equivalent to that of graphene: the staggered sublattice
	potential and the spin-dependent next-nearest-neighbour hopping remain
	finite, even though their competing contributions close the static gap.
	The comparison therefore concerns two distinct gapless reference systems,
	not two realizations of the same Hamiltonian. A vanishing static gap alone
	is thus insufficient to determine the magnitude of the field-induced
	quasienergy gap.
	
	
	Notice that for hBN the induced gap change with the external field polarization, see Table~\ref{tab:floquet_gap_linear_xy}. The two gaps are comparable in size but occur at different combinations of \(E_0\) and \(\omega\), indicating that the coupling to the Floquet spectrum depends on the field direction as well as on the drive strength.
	Notice that even for the highly symmetric graphene lattice, the induced quasienergy
	gap depends on the polarization direction and on the resonance condition set
	by \(\omega\), since $x$ and $y$ are not directly related by symmetry operations.
	\subsection{Topological transition}
\begin{figure}[H]
		\centering
		\begin{minipage}[b]{0.47\linewidth}
			\centering
			\includegraphics[width=\linewidth]
			{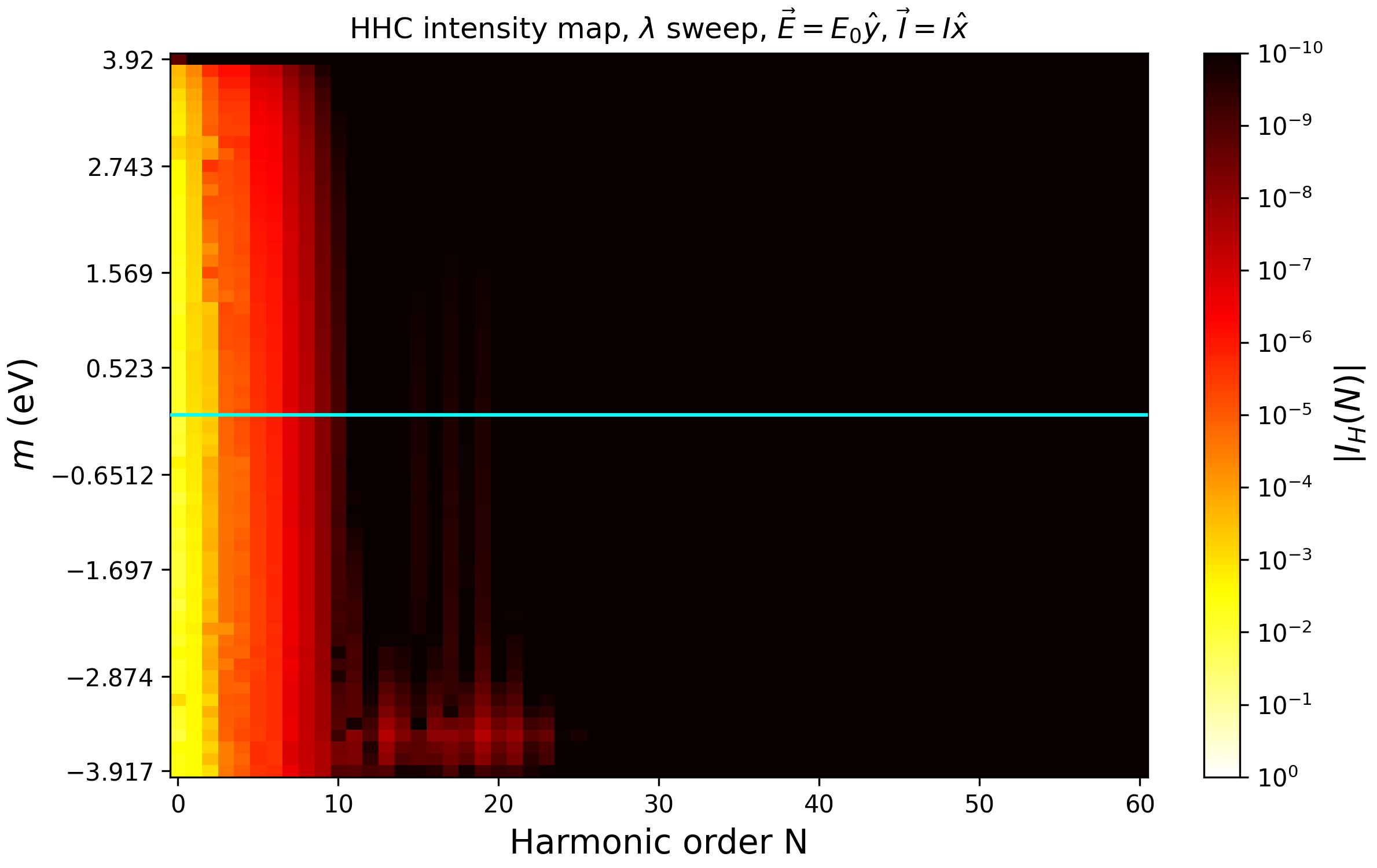}
			\par\smallskip
			(a) \(x\)-polarised drive: \(J_x\).
		\end{minipage}
		\hfill
		\begin{minipage}[b]{0.47\linewidth}
			\centering
			\includegraphics[width=\linewidth]
			{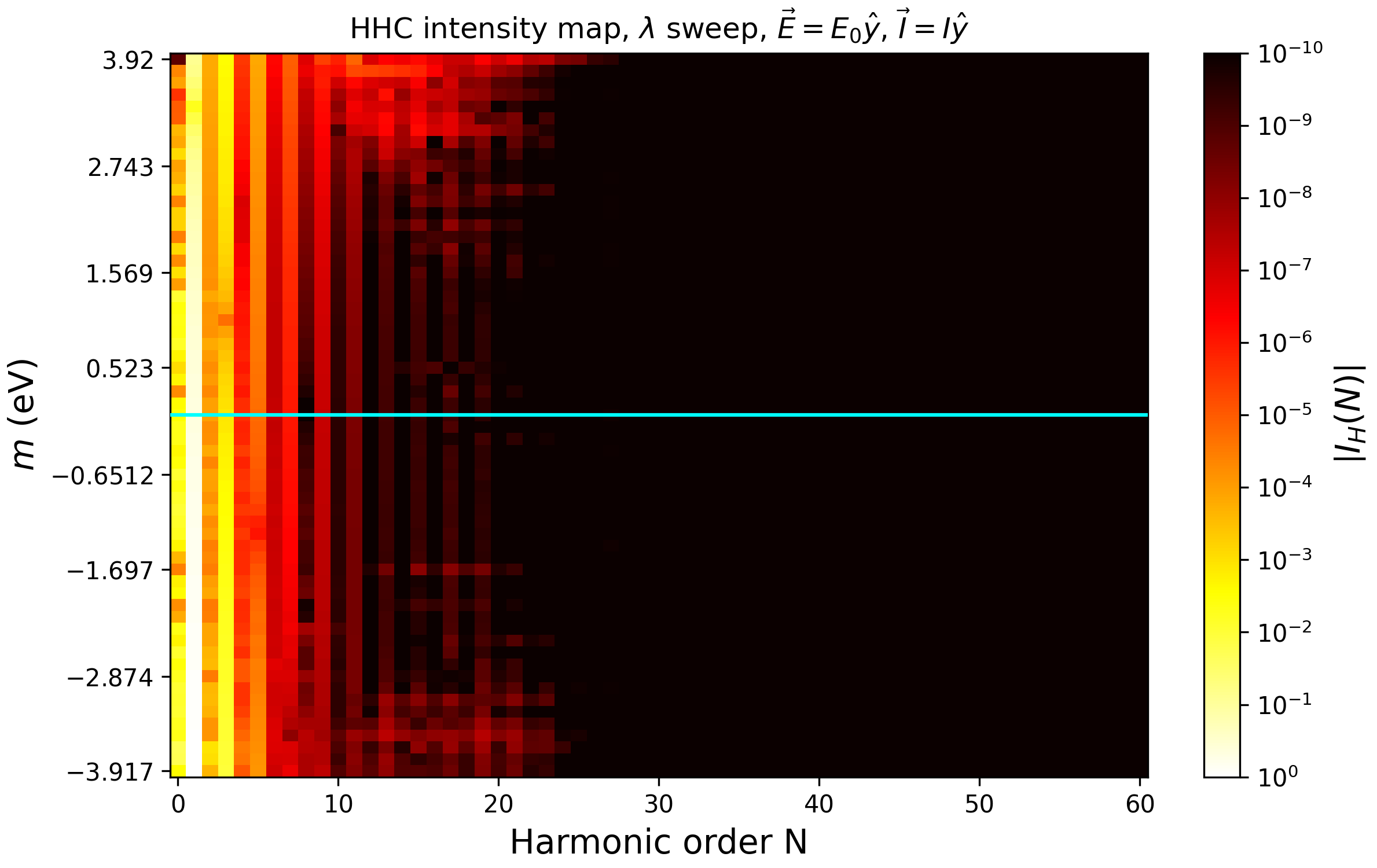}
			\par\smallskip
			(b) \(y\)-polarised drive: \(J_y\).
		\end{minipage}
		\caption{
			Longitudinal HHC spectra as functions of the effective
			Semenoff-mass parameter. The driving field is linearly polarised
			along (a) the \(x\)-direction and (b) the \(y\)-direction.
		}
		\label{fig:hbn_lambda_longitudinal_hhc}
	\end{figure}

	We now vary \(\lambda\) across the critical coupling \(\lambda^\star\) in
	the hBN-based model. The established phase classification distinguishes
	the following three regimes ~\cite{mosquera_topological_2024}:
	\begin{enumerate}
		\item $\mathbf{\lambda < \lambda^\star :}$ Trivial phase
		\item $\mathbf{\lambda = \lambda^\star :}$ Transition phase
		\item $\mathbf{\lambda > \lambda^\star :}$ Topological phase
	\end{enumerate} 
	Following the topological-transition setup of Mosquera \emph{et al.}, which uses helicity-sensitive ultrafast spectra to identify a Chern-insulator transition, we use the parameters ~\cite{mosquera_topological_2024}: $E_0 = 8.68\times10^8$ V/m electric field amplitude;$\omega = 0.4$ eV photon energy; $E_g = 3.92$ energy gap and $\gamma = 0.0~\mathrm{eV}\cdot\text{\AA}$.
	The calculations presented here use sinusoidal vector potentials with linear
	polarisation along either the \(x\)- or \(y\)-direction:
	\begin{align}
		\vec{A}_x(t)
		&=
		\frac{E_0}{\omega}\sin(\omega t)\,\hat{x},
		& \text{(linear along \(x\))}, \\
		\vec{A}_y(t)
		&=
		\frac{E_0}{\omega}\sin(\omega t)\,\hat{y},
		& \text{(linear along \(y\))}.
	\end{align}
	Then the effective Semenoff mass is parametrised as
	\begin{equation}
		m(\lambda)
		=
		2Q
		\left(
		1-\frac{\lambda}{\lambda^\star}
		\right)
		=
		E_g
		\left(
		1-\frac{\lambda}{\lambda^\star}
		\right),
		\qquad
		E_g=2Q,
	\end{equation}
	where \(\lambda^\star\) denotes the critical coupling at which the
	effective mass vanishes.\\
	Figure~\ref{fig:hbn_lambda_longitudinal_hhc} shows the longitudinal
	high-harmonic current (HHC) spectra for driving fields linearly polarised
	along the \(x\)- and \(y\)-directions. In each case, the current component
	parallel to the driving field is shown. For the parameters and linear
	polarisations considered here, the longitudinal HHC spectra vary only weakly
	across the topological transition. This observation does not imply that
	topology has no influence on HHG. Rather, it is consistent with the caution
	raised by Neufeld \emph{et al.}~\cite{neufeld2023there} that HHG observables
	need not provide universal or unambiguous signatures of topological phases.
	
\section{Code structure} \label{usage}
In this section we briefly present the code structure and how it is organized in the different files.

\begin{figure}[H]
\centering
\begin{tikzpicture}[
    file/.style={draw, rounded corners, align=center, minimum width=3.0cm, minimum height=0.8cm},
    arrow/.style={->, >=Stealth, thick},
    node distance=1.0cm and 1.8cm
]

\node[file] (init)    {initializer.jl};
\node[file, below=of init]    (ham)     {Ham.jl};
\node[file, below=of ham]     (current) {current.jl};
\node[file, below=of current] (plots)   {Plots.jl};
\node[file, below=of plots]   (main)    {main\_F.jl};

\node[file, left=2.8cm of init, yshift=1.5cm]  (tools)     {tools.jl};
\node[file, left=2.8cm of init, yshift=0.5cm]  (lattice)   {Lattice.jl};
\node[file, left=2.8cm of init, yshift=-0.5cm] (constants) {constants.jl};
\node[file, left=2.8cm of init, yshift=-1.5cm] (unitcell)  {unit\_cell.jl};

\draw[arrow] (tools) -- (init);
\draw[arrow] (lattice) -- (init);
\draw[arrow] (constants) -- (init);
\draw[arrow] (unitcell) -- (init);

\draw[arrow] (init) -- (ham);
\draw[arrow] (ham) -- (current);
\draw[arrow] (current) -- (plots);
\draw[arrow] (plots) -- (main);

\end{tikzpicture}
\caption{Module/dependency flow for the Julia code used in this work.}
\label{fig:code-flow}
\end{figure}
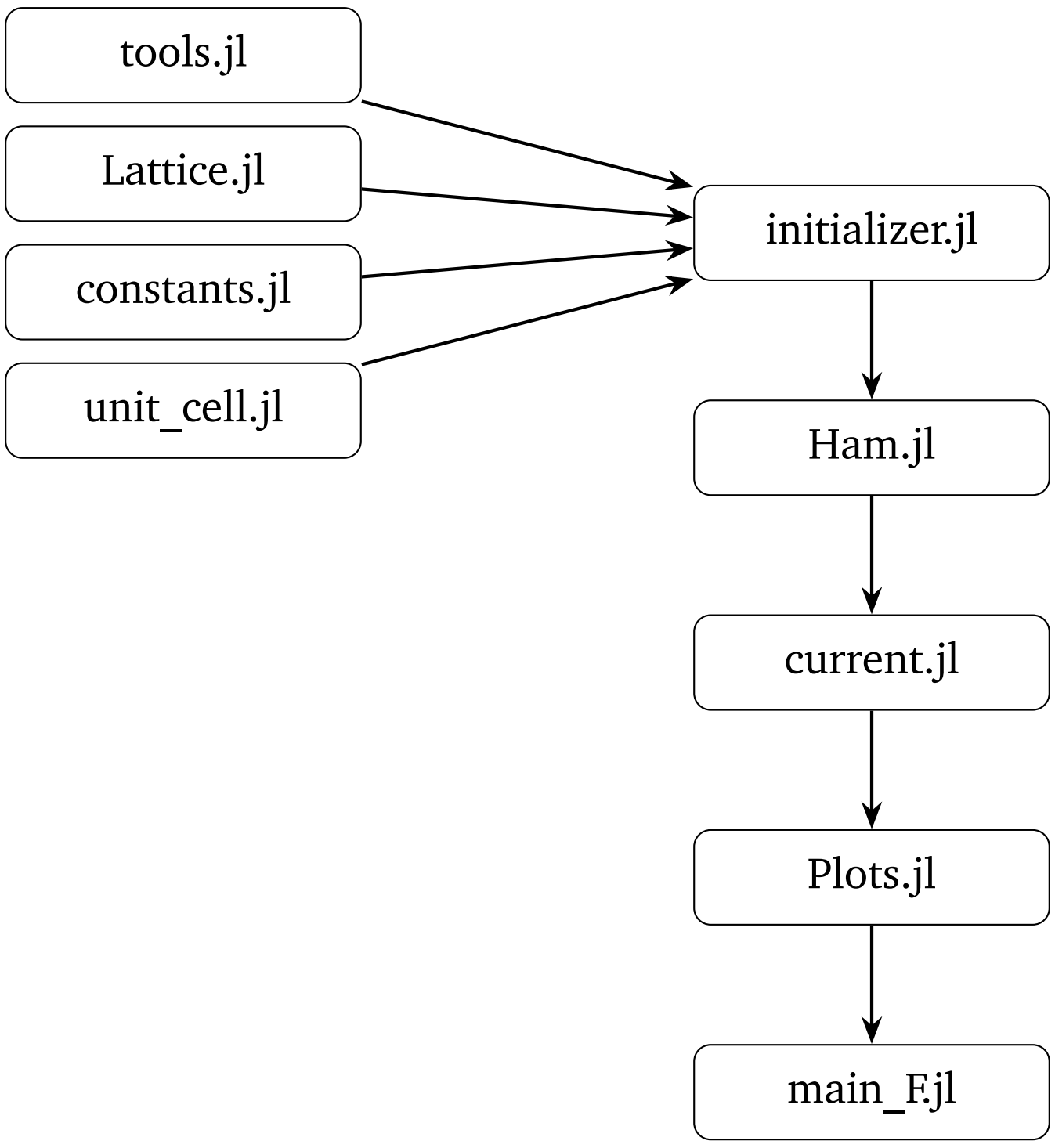

The \texttt{tools.jl} module provides interface utilities and common numerical routines. The \texttt{Lattice.jl} module analyzes the direct and reciprocal lattice vectors, computes the unit-cell and Brillouin-zone volumes, and defines the \(\mathbf{k}\)-point grid. The \texttt{constants.jl} module contains the conversion factors for the supported unit systems: SI, custom, and atomic units, with custom units selected by default. The \texttt{unit\_cell.jl} module reads the coordinate file (\texttt{.coo}), identifies nearest-neighbour and next-nearest-neighbour bonds and their chiralities, and constructs \(\mathbb{H}_{\vec{R};ij}\) used in Eq.~\eqref{eq:static_tb_hamiltonian}. The \texttt{initializer.jl} module defines the simulation parameters. The \texttt{Ham.jl} module evaluates the Bessel functions, constructs the tight-binding and Floquet Hamiltonians, and computes their eigenvalues and eigenvectors. The \texttt{current.jl} module evaluates the current operator and the high-harmonic current (HHC), while \texttt{Plots.jl} generates the corresponding figures. Finally, \texttt{main\_F.jl} connects these modules and serves as the main executable.
To run the Julia code, replace the placeholders in the command below with
the desired system name and numerical parameter values. The input units and
polarisation codes are listed below; units are not included in the command-line
arguments.

\texttt{julia --project main\_F.jl -t auto -I \textbf{<system\_name>} \\ -Q \textbf{<Energy gap>}\\ -E \textbf{<Amplitude of Electric Field>} \\ -S \textbf{<SOC coupling>} \\ -R \textbf{<Rashba coupling>} \\ -P \textbf{<Polarisation of the Electric Field>} }

\begin{center}
\begin{tabular}{lll}
\toprule
Option & Parameter & Input unit \\
\midrule
\texttt{-I} & System name & --- \\
\texttt{-Q} & Energy gap & eV \\
\texttt{-E} & Electric-field amplitude & V/m \\
\texttt{-S} & Spin-orbit coupling & eV \\
\texttt{-R} & Rashba coupling & eV \\
\texttt{-P} & Polarisation code & Dimensionless \\
\bottomrule
\end{tabular}
\end{center}

The polarisation code is \texttt{0} for linear polarisation along the
\(x\)-axis, \texttt{2} for linear polarisation along the \(y\)-axis,
\texttt{1} for right-circular polarisation, and \texttt{-1} for left-circular
polarisation. The corresponding vector potentials are given in
Appendix~\ref{app:fft}, Eq.~\eqref{eq:code_polarisation_options}.
For example, \texttt{-E 1.0e9 -P 0} specifies an electric-field amplitude of
\(10^9~\mathrm{V/m}\) (\(1~\mathrm{V/nm}\)) with linear polarisation along
the \(x\)-axis.

\section{Conclusions}\label{conclusions}
	
	We have rewritten the Floquet treatment of high-harmonic generation as a
	compact Fourier-Sambe formalism for driven tight-binding Hamiltonians. The
	method starts from the Fourier components of the Peierls-driven Hamiltonian,
	constructs the Sambe-space eigenproblem, and evaluates the long-time harmonic
	current from the diagonal Floquet populations. Off-diagonal Floquet coherences
	generate quasienergy beat frequencies and contribute to pure integer
	harmonics only under a resonance condition.
	
	Applied to honeycomb models of hBN and graphene, the method shows clear
	polarization selection in the high-harmonic current. In hBN, longitudinal
	response to \(x\)-polarized fields is dominated by odd harmonics, while the
	transverse response highlights even harmonics. We have shown how, using our code, 
	it is possible to study the HHG response as a function of the Hamiltonian parameters, such as, 
	for example, the hopping parameter. 
	Finally, in the Kane-Male model, we have shown just how the HHG response could be insensitive to a topological transition.\\
	Our future work will focus on optimizing the Floquet Hamiltonian construction for high-performance computing architectures 
	and expanding the package to treat complex 3D spin-orbit coupled systems, including an interface to Wannier-tight binding Hamiltonians.

\section*{Acknowledgments}
		C.A. and A.P. acknowledge B. Demoulin and A. Saul for the management of the computer cluster \emph{Claudia}. C.A., A.M. and A.P. acknowledge support from the Amidex foundation through the project INDIGENA.  C.A. acknowledges funding from European Research Council MSCA-ITN TIMES under grant agreement 101118915.  
\appendix
\section{Floquet Hamiltonian and Fast Fourier Transform (FFT)}
\label{app:fft}
The main-text derivation uses monochromatic, linearly polarised fields, whereas
the numerical implementation allows a more general laser profile. The full
vector potential is written as
\begin{equation}
	\vec{A}(t)
	=
	e_{\mathrm{env}}(t)\vec{A}_P(t),
	\label{eq:full_vector_potential}
\end{equation}
where the coding parameter \(P\) selects one of the predefined polarisation
states,
\begin{equation}
	\vec{A}_P(t)
	=
	-\frac{E_0}{\omega}
	\begin{cases}
		\sin(\omega t)\,\hat{x},
		& P=0 \quad (x\text{-linear}), \\[3pt]
		\sin(\omega t)\,\hat{y},
		& P=2 \quad (y\text{-linear}), \\[3pt]
		\sin(\omega t)\,\hat{x}+\cos(\omega t)\,\hat{y},
		& P=+1 \quad (\text{right-circular}), \\[3pt]
		\sin(\omega t)\,\hat{x}-\cos(\omega t)\,\hat{y},
		& P=-1 \quad (\text{left-circular}).
	\end{cases}
	\label{eq:code_polarisation_options}
\end{equation}
For a continuous monochromatic drive, \(e_{\mathrm{env}}(t)=1\). Alternatively,
an envelope satisfying \(e_{\mathrm{env}}(0)=0\) ensures that
\(\vec{A}(0)=\vec{0}\); a \(\sin^2\) envelope is one possible choice. Both the
polarisation components and the envelope can be replaced by user-defined
functions on the numerical time grid. The FFT implementation can therefore
accommodate arbitrary polarisation states and pulse envelopes, providing a
flexible description of a broad range of laser profiles.

For a monochromatic drive and simple tight-binding models, the Fourier
components of the Hamiltonian can sometimes be written analytically in terms
of Bessel functions, as in the collinear-dimer benchmark of
Ref.~\cite{ikeda_floquet-theoretical_2018}. In the code, however, we also want to treat more general
periodic driving protocols and Hamiltonians for which
closed-form Fourier coefficients are inconvenient. We therefore evaluate the
Fourier components numerically on a uniform time grid using the Fast Fourier
Transform implementation FFTW~\cite{frigo1998fftw}.

For each momentum \(\mathbf{k}\), the Hamiltonian is sampled at
\(N_t\) points over one period,
\begin{equation}
	t_j=\frac{jT}{N_t},
	\qquad
	j=0,\ldots,N_t-1,
	\qquad
	T=\frac{2\pi}{\omega_0}.
\end{equation}
The discrete Fourier coefficient used in the Sambe Hamiltonian is
\begin{equation}
	H^{(\ell)}_{\mu\nu}(\mathbf{k})
	\simeq
	\frac{1}{N_t}
	\sum_{j=0}^{N_t-1}
	\ee^{-\ii\ell\omega_0 t_j}
	H_{\mu\nu}(t_j,\mathbf{k})
	\equiv
	\frac{1}{N_t}
	\operatorname{FFT}_{\ell}
	\left[H_{\mu\nu}(t_j,\mathbf{k})\right],
	\label{eq:fft_hamiltonian_component}
\end{equation}
where only the harmonics \(\ell\in[-\NF,\NF]\) are retained. These coefficients
are inserted into the truncated Sambe matrix,
\begin{equation}
	\left[
	H^F(\mathbf{k})
	\right]_{m\mu,n\nu}
	=
	H^{(m-n)}_{\mu\nu}(\mathbf{k})
	+
	n\hbar\omega_0
	\delta_{mn}\delta_{\mu\nu},
	\label{eq:fft_sambe_hamiltonian}
\end{equation}
which is then diagonalized to obtain the quasienergies and Fourier components
of the Floquet modes.

After constructing the Floquet state in Eq.~\eqref{eq:current_operator}, the
current operator is evaluated from the Peierls-shifted Hamiltonian,
\begin{equation}
	\hat{\mathcal I}_a(t,\mathbf{k})
	=
	-\frac{\partial H(t,\mathbf{k})}{\partial A_a(t)}
	=
	-\frac{q_e}{\hbar}
	\left.
	\frac{\partial H_0(\mathbf{k}')}{\partial k'_a}
	\right|_{\mathbf{k}'=\mathbf{k}+q_e\mathbf{A}(t)/\hbar}.
	\label{eq:fft_current_operator}
\end{equation}
When an analytic derivative is not available, the code uses a fourth-order
central finite difference along direction \(a\),
\begin{equation}
	\frac{\partial H_0(\mathbf{k})}{\partial k_a}
	\simeq
	\frac{
	H_0(\mathbf{k}-2\Delta k_a\hat{\mathbf e}_a)
	-8H_0(\mathbf{k}-\Delta k_a\hat{\mathbf e}_a)
	+8H_0(\mathbf{k}+\Delta k_a\hat{\mathbf e}_a)
	-H_0(\mathbf{k}+2\Delta k_a\hat{\mathbf e}_a)
	}{12\Delta k_a}
	+
	\mathcal{O}(\Delta k_a^4).
	\label{eq:fourth_order_current_derivative}
\end{equation}

The time-dependent current component is then computed as
\begin{equation}
	I_a(t)
	=
	\sum_{\mathbf{k}}
	\mel{\psi(t,\mathbf{k})}
	{\hat{\mathcal I}_a(t,\mathbf{k})}
	{\psi(t,\mathbf{k})}.
	\label{eq:fft_total_current}
\end{equation}
For the strictly harmonic part of the response we retain the diagonal Floquet
population contribution, consistent with Eq.~\eqref{eq:harmonic_current},
\begin{equation}
	I^H_a(t)
	=
	\sum_{\mathbf{k}}
	\sum_u
	\abs{\gamma_u(\mathbf{k})}^2
	\left[
	\chi^\dagger(t,\mathbf{k})
	\hat{\mathcal I}_a(t,\mathbf{k})
	\chi(t,\mathbf{k})
	\right]_{uu}.
	\label{eq:fft_harmonic_current_time}
\end{equation}
Finally, the high-harmonic current spectrum is obtained by a second FFT over
the sampled current,
\begin{equation}
	I^H_a(N)
	\simeq
	\frac{1}{N_t}
	\sum_{j=0}^{N_t-1}
	\ee^{-\ii N\omega_0t_j}
	I^H_a(t_j)
	\equiv
	\frac{1}{N_t}
	\operatorname{FFT}_{N}
	\left[I^H_a(t_j)\right].
	\label{eq:fft_hhc_spectrum}
\end{equation}
This FFT route uses the same Floquet-Sambe structure as the analytic
Fourier-Bessel construction, but replaces closed-form Fourier coefficients by
their discrete numerical counterparts.
	\bibliography{Floquet,hhg}
\nolinenumbers
\end{document}